\documentclass[12pt, preprint]{aastex}

\usepackage{amssymb}
\usepackage{amsmath}
\usepackage{epsfig}
\usepackage{graphicx}
\usepackage{natbib}
\usepackage{multirow}
\usepackage{rotating}
\usepackage{longtable}
\usepackage{verbatim}
\usepackage{url}
\usepackage{hyperref}
\usepackage{caption2}

\begin{document}

\title{
A Detailed Model Grid for Solid Planets from $0.1$ through $100$ Earth Masses
}
\author{Li Zeng, Dimitar Sasselov
}
\affil{Astronomy Department, Harvard University, Cambridge, MA 02138}
\email{lzeng@cfa.harvard.edu} 
\email{dsasselov@cfa.harvard.edu}

\begin{abstract}
This paper describes a new grid for the mass-radius relation of $3$-layer exoplanets within the mass range of $0.1$ through $100$ $M_{\oplus}$. The $3$ layers are: Fe ($\epsilon$-phase of iron), MgSiO${}_{3}$ (including both the perovskite phase, post-perovskite phase, and its dissociation at ultra-high pressures), and H${}_{2}$O (including Ices Ih, III, V, VI, VII, X, and the superionic phase along the melting curve). We discuss the current state of knowledge about the equations of state (EOS) that influence these calculations and the improvements used in the new grid. For the $2$-layer model, we demonstrate the utility of contours on the mass-radius diagrams. Given the mass and radius input, these contours can be used to quickly determine the important physical properties of a planet including its p0 (central pressure), p1/p0 (core-mantle boundary pressure over central pressure), CMF (core mass fraction) or CRF (core radius fraction). For the $3$-layer model, a curve segment on the ternary diagram represents all possible relative mass proportions of the $3$ layers for a given mass-radius input. These ternary diagrams are tabulated into Table~\ref{Table3} with the intent to make comparison to observations easier. How the presence of Fe in the mantle affects the mass-radius relations is also discussed in a separate section. A dynamic and interactive tool to characterize and illustrate the interior structure of exoplanets built upon models in this paper is available on the website:~\url{http://www.cfa.harvard.edu/~lzeng}. 

\end{abstract}
\keywords{exoplanet, Super-Earth, interior structure modeling, Kepler Mission}

\section{Introduction}
The transit method of exoplanet discovery has produced a small, but well constrained, sample of exoplanets that are unambiguously solid in terms of interior bulk composition. We call solid planets the ones that possess no H and He envelopes and/or atmospheres, i.e. their bulk radius is determined by elements (and their mineral phases) heavier than H and He. Such solid exoplanets are Kepler-10b~\citep[][]{Batalha:2011}, CoRoT-7b~\citep[][]{Queloz:2009, Hatzes:2011}, Kepler-36b~\citep[][]{Carter:2012} as well as - most likely: Kepler-20b,e,f; Kepler-18b, and 55 Cnc e in which the solid material could include high-pressure water ice (see references to Fig.~\ref{MRplot} in~\ref{4.2}). 

There is an increased interest to compare their observed parameters to current models of interior planetary structure. The models, and their use of approximations and EOS, have evolved since 2005~\citep[][]{Valencia:2006, Fortney:2007, Grasset:2009, Seager:2007}, mainly because the more massive solid exoplanets (called Super-Earths) have interior pressures that are far in excess of Earth's model, bringing about corresponding gaps in our knowledge of mineral phases and their EOS~\citep[see a recent review by][]{Seager:2010}. Here we compute a new grid of models in order to aid current comparisons to observed exoplanets on the mass-radius diagram. As in previous such grids, we assume the main constituents inside the planets to be differentiated and model them as layers in one dimension. 

The first part of this paper aims to solve the $2$-layer exoplanet model. The $2$-layer model reveals the underlying physics of planetary interior more intuitively, for which we consider 3 scenarios: Fe-MgSiO${}_{3}$ or Fe-H${}_{2}$O or MgSiO${}_{3}$-H${}_{2}$O planet. 

The current observations generally measure the radius of an exoplanet through transits and the mass through Doppler shift measurement of the host star. For each assumption of core and mantle compositions, given the mass and radius input, the $2$-layer exoplanet model can be solved uniquely. It is a unique solution of radial dependence of interior pressure and density. As a result, all the characteristic physical quantities, such as the pressure at the center (p0), the pressure at core-mantle boundary (p1), the core mass fraction (CMF), and the core radius fraction (CRF) naturally fall out from this model. These quantities can be quickly determined by invoking the mass-radius contours. 

The next part of this paper (Fig.~\ref{MRplot}) compares some known exoplanets to the mass-radius curves of 6 different $2$-layer exoplanet models: pure-Fe, $50\%$-Fe \& $50\%$-MgSiO${}_{3}$, pure MgSiO${}_{3}$, $50\%$-MgSiO${}_{3}$ \& $50\%$-H${}_{2}$O, $25\%$-MgSiO${}_{3}$ \& $75\%$-H${}_{2}$O, and pure H${}_{2}$O. These percentages are in mass fractions. The data of these six curves are available in Table~\ref{Table1}. 

Up to now, a standard assumption has been that the planet interior is fully differentiated into layers: all the Fe is in the core and all the MgSiO${}_{3}$ is in the mantle. In section~\ref{nondiff}, we will change this assumption and discuss how the presence of Fe in the mantle affects the mass-radius relation. 

The final part of this paper calculates the $3$-layer differentiated exoplanet model. Given the mass and radius input, the solution for the $3$-layer model is non-unique (degenerate), thus a curve on the ternary diagram is needed to represent the set of all solutions (see Fig.~\ref{ternary}). This curve can be obtained by solving differential equations with iterative methods. The ensemble of solutions is tabulated (Table~\ref{Table3}), from which users may interpolate to determine planet composition in $3$-layer model. A dynamic and interactive tool to characterize and illustrate the interior structure of exoplanets built upon Table~\ref{Table3} and other models in this paper is available on the website \url{http://www.cfa.harvard.edu/~lzeng}. 

The methods described in this paper can be used to fast characterize the interior structure of exoplanets. 

\section{Method}
Spherical symmetry is assumed in all the models. The interior of a planet is assumed to be fully differentiated into layers in the first part of the paper. 
The $2$-layer model consists of a core and a mantle. The $3$-layer model consists of a core, a mantle and another layer on top of the mantle. The interior structure is determined by solving the following two differential equations: 

\begin{equation}
\frac{dr}{dm}=\frac{1}{4 \pi \rho r^2}
\label{dfeq1}
\end{equation}

\begin{equation}
\frac{dp}{dm}=-\frac{G m}{4 \pi r^4}
\label{dfeq2}
\end{equation}

The two equations are similar to the ones in~\cite{Zeng_Seager:2008}. However, contrary to the common choice of radius r as the independent variable, the interior mass m is chosen, which is the total mass inside shell radius r, as the independent variable. So the solution is given as r(m) (interior radius r as a dependent function of interior mass m), p(m) (pressure as a dependent function of interior mass m), and $\rho$(m) (density as a dependent function of interior mass m). 

The two differential equations are solved with the EOS of the constituent materials as inputs:

\begin{equation}
\rho=\rho(p,T)
\label{EOS}
\end{equation}

The EOS is a material property, which describes the material's density as a function of pressure and temperature. The EOS could be obtained both theoretically and experimentally. 
Theoretically, the EOS could be calculated by Quantum-Mechanical Molecular Dynamics Ab Initio Simulation such as the VASP (Vienna Ab initio Simulation Package)~\citep[][]{Kresse:1993, Kresse:1994, Kresse:1996, French:2009}. 
Experimentally, the EOS could be determined by high-pressure compression experiment such as the DAC (Diamond Anvil Cell) experiment, or shock wave experiment like the implosion experiment by the Sandia Z-machine~\citep{Yu:2011}. 
The temperature effect on density is secondary compared to the pressure effect~\citep{Valencia:2006, Valencia:2007a}. 
Therefore, we can safely ignore the temperature dependence of those higher density materials (Fe and MgSiO${}_{3}$) for which the temperature effect is weaker, or we can implicitly include a pre-assumed pressure-temperature (p-T) relation (for H${}_{2}$O it is the melting curve) so as to reduce the EOS to a simpler single-variable form: 

\begin{equation}
\rho=\rho(p)
\label{EOS2}
\end{equation}

To solve the set of equations mentioned above, appropriate boundaries conditions are given as:

\begin{itemize}
\item{p0:} the pressure at the center of the planet
\item{p1:} the pressure at the first layer interface (the core-mantle boundary)
\item {p2:} the pressure at the second layer interface (only needed for $3$-layer model)
\item {p${}_{surface}$}: the pressure at the surface of the planet (set to 1 bar ($10^5 Pa$))
\end{itemize}

\section{EOS of Fe, MgSiO${}_{3}$ and H${}_{2}$O}
The 3 layers that we consider for the planet interior are Fe, MgSiO${}_{3}$ and H${}_{2}$O.
Their detailed EOS are described as follows: 

\subsection{Fe}
We model the core of a solid exoplanet after the Earth's iron core, except that in our model we ignore the presence of all other elements such as Nickel (Ni) and light elements such as Sulfur (S) and Oxygen (O) in the core. 
As pointed out by~\citet{Valencia:2010}, above $100$ GPa, the iron is mostly in the hexagonal closed packed $\epsilon$ phase. Therefore, we use the Fe-EOS by~\citet{Seager:2007}. Below $2.09*10^{4}$ GPa, it is a Vinet~\citep{Vinet:1987, Vinet:1989} formula fit to the experimental data of $\epsilon$-iron at p$\leq330$ GPa by~\citet{Anderson:2001}. Above $2.09*10^{4}$ GPa, it makes smooth transition to the Thomas-Fermi-Dirac (TFD) EOS~\citep{Seager:2007}. A smooth transition is assumed because there is no experimental data available in this ultrahigh-pressure regime. 

The central pressure could reach $250$ TPa (terapascal, i.e., $10^{12}$ Pa) in the most massive planet considered in this paper. However, the EOS of Fe above $400$ GPa is beyond the current reach of experiment and thus largely unknown. Therefore, our best approximation here is to extend the currently available $\epsilon$-iron EOS to higher pressures and connect it to the TFD EOS.

\begin{figure}[htbp]
\begin{center}
\includegraphics[scale=0.9]{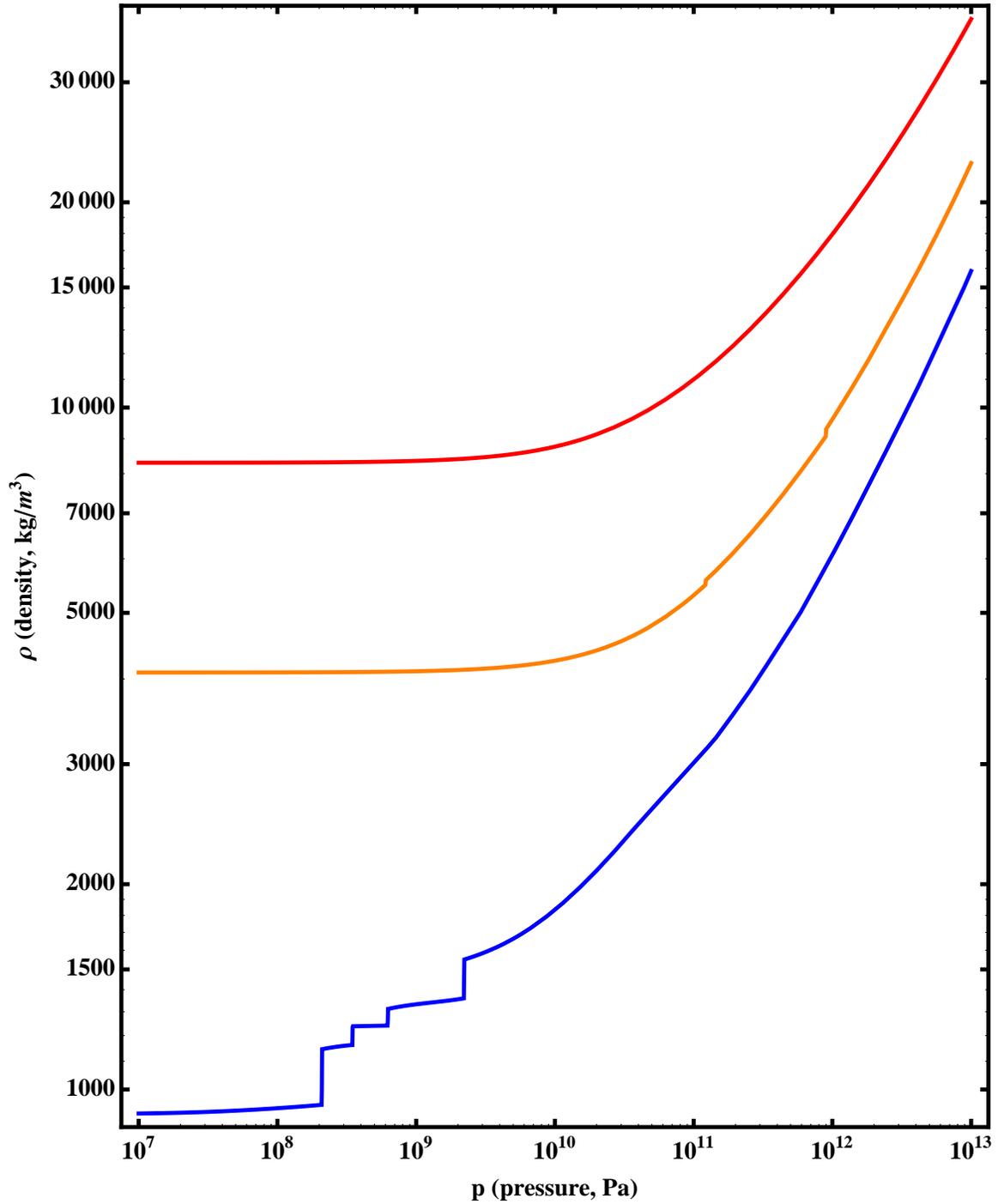}
\end{center}
\caption{$p-\rho$ EOS of Fe ($\epsilon$-phase of iron, red curve), MgSiO${}_{3}$ (perovskite phase, post-perovskite phase and its high-pressure derivatives, orange curve), and H${}_{2}$O (Ice Ih, Ice III, Ice V, Ice VI, Ice VII, Ice X, and superionic phase along its melting curve (solid-liquid phase boundary))}
\label{eosplot}
\end{figure}

The EOS of Fe is shown in Fig.~\ref{eosplot} as the upper curve (red curve). 

\subsection{MgSiO${}_{3}$}
We model the silicate layer of a solid exoplanet using the Earth's mantle as a proxy. The FeO-free Earth's mantle with Mg/Si=1.07 would consist of mainly enstatite (MgSiO${}_{3}$) or its high-pressure polymorphs and, depending upon pressure, small amounts of either forsterite and its high-pressure polymorphs (Mg${}_{2}$SiO${}_{4}$) or periclase (MgO)~\citep[e.g.][]{Bina:2003}.

The olivine polymorphs as well as lower-pressure enstatite and majorite (MgSiO${}_{3}$ with garnet structure), are not stable above 27 GPa. At higher pressures, the system would consist of MgSiO${}_{3}$-perovskite (pv) and periclase or their higher-pressure polymorphs~\citep{Stixrude:2011, Bina:2003}. Given the high pressures at the H${}_{2}$O-silicate boundary usually exceeding 27 GPa, we can safely ignore olivine and lower-pressure pyroxene polymorphs. For the sake of simplicity, we also ignore periclase, which would contribute only 7 at.$\%$ to the silicate mantle mineralogical composition. There are also small amounts of other elements such as Aluminum (Al), Calcium (Ca), and Sodium (Na) present in Earth's mantle~\citep{Sotin:2007}. For simplicity, we  neglect them and thus the phases containing them are not included in our model. We also do not consider SiC because carbon-rich planets might form under very rare circumstances, and are probably not common. 

Some fraction of Fe can be incorporated into the minerals of the silicate mantle which could then have the general formula as (Mg,Fe)SiO${}_{3}$. For now, we simply assume all the Fe is in the core and all the Mg is in the mantle in the form of MgSiO${}_{3}$-perovskite and/or its high-pressure polymorphs. So the planet is fully differentiated. In a later section, we will discuss how the addition of Fe to the mantle can affect mass-radius relation and compare the differences between differentiated and undifferentiated as well as reduced and oxidized planets in Section~\ref{nondiff}. 

We first consider the perovskite (pv) and post-perovskite (ppv) phases of pure MgSiO${}_{3}$. MgSiO${}_{3}$ perovskite (pv) is believed to be the major constituent of the Earth mantle. It makes transition into the post-perovskite (ppv) phase at roughly $120$ GPa and 2500 K (corresponding to a depth of 2600 kilometers in Earth)~\citep{Hirose:2010}. The ppv phase was discovered experimentally in 2004~\citep[by][]{Murakami:2004} and was also theoretically predicted in the same year~\citep[by][]{Oganov:2004}. The ppv is about 1.5$\%$ denser than the pv phase~\citep{Caracas:2008, Hirose:2010}. This 1.5$\%$ density jump resulting from the pv-to-ppv phase transition can be clearly seen as the first density jump of the MgSiO${}_{3}$ EOS curve shown in Fig.~\ref{eosplot}. Both the MgSiO${}_{3}$ pv EOS and MgSiO${}_{3}$ ppv EOS are taken from~\citet{Caracas:2008}. The transition pressure is determined to be $122$ GPa for pure MgSiO${}_{3}$ according to~\citet{Spera:2006}. 

Beyond $0.90$ TPa, MgSiO${}_{3}$ ppv undergoes a two-stage dissociation process predicted from the first-principle calculations by~\citet{Umemoto:2011}. MgSiO${}_{3}$ ppv first dissociates into CsCl-type MgO and P2${}_{1}$c-type MgSi${}_{2}$O${}_{5}$ at the pressure of $0.90$ TPa and later into CsCl-type MgO and Fe${}_{2}$P-type SiO${}_{2}$ at pressures higher than $2.10$ TPa. The EOS of CsCl-type MgO, P2${}_{1}$c-type MgSi${}_{2}$O${}_{5}$, and Fe${}_{2}$P-type SiO${}_{2}$ are adopted from~\citet{Umemoto:2011, Wu:2011}. Therefore, there are two density jumps at the dissociation pressures of $0.90$ TPa and $2.10$ TPa. The first one can be seen clearly in Fig.~\ref{eosplot}. The second one cannot be seen in Fig.~\ref{eosplot} since it is too small, but it surely exists. 

Since~\citet{Umemoto:2011}'s EOS calculation is until $4.90$ TPa, beyond $4.90$ TPa, a modified version of the EOS by~\citet{Seager:2007} is used to smoothly connect to the TFD EOS. TFD EOS assumes electrons in a slowly varying potential with a density-dependent correlation energy term that describes the interactions among electrons. It is therefore insensitive to any crystal structure or arrangements of atoms and it becomes asymptotically more accurate at higher pressure. Thus, the TFD EOS of MgSiO${}_{3}$ would be no different from the TFD EOS of MgO plus SiO${}_{2}$ as long as the types and numbers of atoms in the calculation are the same. So it is safe to use the TFD EOS of MgSiO${}_{3}$ as an approximation of the EOS of MgO and SiO${}_{2}$ mixture beyond $4.90$ TPa here. 

\citet{Seager:2007}'s EOS is calculated from the method of~\citet{Salpeter:1967} above $1.35*10^{4}$ GPa. Below $1.35*10^{4}$ GPa, it is a smooth connection to TFD EOS from the fourth-order Birch-Murbaghan Equation of State (BME)~\citep[see][]{Birch:1947, Poirier:2000} fit to the parameters of MgSiO${}_{3}$ pv obtained by \textit{Ab initio} lattice dynamics simulation of~\citet{Karki:2000}. 

~\citeauthor{Seager:2007}'s EOS is slightly modified to avoid any artificial density jump when connected with~\citeauthor{Umemoto:2011}'s EOS at $4.90$ TPa. At $4.90$ TPa, the ratio of the density $\rho$ between~\citeauthor{Umemoto:2011}'s EOS and~\citeauthor{Seager:2007}'s EOS is $1.04437$. This ratio is multiplied to the original~\citeauthor{Seager:2007}'s EOS density $\rho$ to produce the actual EOS used in our calculation for p$>4.90$ TPa. 
A smooth transition is assumed because no experimental data is available in this ultrahigh-pressure regime. This assumption does not affect our low or medium mass planet models, since only the most massive planets in our model could reach this ultrahigh pressure in their MgSiO${}_{3}$ part. 

The EOS of MgSiO${}_{3}$ is shown in Fig.~\ref{eosplot} as the middle curve (orange curve). 

\subsection{H${}_{2}$O}
The top layer of a planet could consist of various phases of H${}_{2}$O. 
Since H${}_{2}$O has a complex phase diagram, and it also has a stronger temperature dependence, thus the temperature effect cannot be ignored. Instead, we follow the solid phases along the melting curve (solid-liquid phase boundary on the p-T plot by~\citet{Chaplin:2012}).
Along the melting curve, the H${}_{2}$O undergoes several phase transitions. Initially, it is Ice Ih at low pressure, then subsequently transforms into Ice III, Ice V, Ice VI, Ice VII, Ice X, and superionic phase~\citep{Chaplin:2012, Choukroun:2007, Dunaeva:2010, French:2009}. 

\subsubsection{Chaplin's EOS}
The solid form of water has very complex phases in the low-pressure and low-temperature regime. These phases are well determined by experiments. Here we adopt the Chaplin's EOS for Ice Ih, Ice III, Ice V, and Ice VI below $2.216$ GPa~\citep[see][]{Chaplin:2012, Choukroun:2007, Dunaeva:2010}. Along the melting curve (the solid-liquid boundary on the p-T diagram), the solid form of water first exists as Ice Ih (Hexagonal Ice) from ambient pressure up to $209.5$ MPa~\citep{Choukroun:2007}. At the triple point of $209.5$ MPa and $251.15$ K~\citep{Choukroun:2007, Robinson:1996}, it transforms into Ice III (Ice-three), whose unit cell forms tetragonal crystals. Ice III exists up to $355.0$ MPa and transforms into a higher-pressure form Ice V (Ice-five) at the triple point of $355.0$ MPa and $256.43$ K~\citep{Choukroun:2007}. Ice V's unit cell forms monoclinic crystals. At the triple point of $618.4$ MPa and $272.73$ K~\citep{Choukroun:2007}, Ice V transforms into yet another higher-pressure form  Ice VI (Ice-six). Ice VI's unit cell forms tetragonal crystals. A single molecule in Ice VI crystal is bonded to four other water molecules. Then at the triple point of $2.216$ GPa and $355$ K~\citep{IAPWS:2011}, Ice VI transforms into Ice VII (Ice-seven). Ice VII has a cubic crystal structure. Ice VII eventually transforms into Ice X (Ice-ten) at the triple point of $47$ GPa and $1000$ K~\citep{Goncharov:2005}. In Ice X, the protons are equally spaced and bonded between the oxygen atoms, where the oxygen atoms are in a body-centered cubic lattice~\citep{Hirsch:1984}. The EOS of Ice X and Ice VII are very similar. For Ice VII (above $2.216$ GPa), we switch to the~\citep*{Frank:2004}'s EOS (FFH2004). 

\subsubsection{FFH2004's EOS}
We adopt FFH2004's EOS of Ice VII for $2.216$ GPa$\leq$p$\leq37.4$ GPa. This EOS is obtained using the Mao-Bell type diamond anvil cell with an external Mo-wire resistance heater. Gold and water are put into the sample chamber and compressed. The diffraction pattern of both H${}_{2}$O and gold are measured by the Energy-Dispersive X-ray Diffraction (EDXD) technique at the Brookhaven National Synchrotron Light Source. The gold here is used as an internal pressure calibrant. The disappearance of the diffraction pattern of the crystal Ice VII is used as the indicator for the solid-liquid boundary (melting curve). The melting curve for Ice VII is determined accurately from $3$ GPa to $60$ GPa and fit to a Simon equation. The molar density ($\rho$ in $mol/cm^3$) of Ice VII as a function of pressure (p in GPa) is given by the following formula in FFH2004:

\begin{equation}
\rho(mol/cm^3)=0.0805+0.0229*(1-exp^{-0.0743*p})+0.1573*(1-exp^{-0.0061*p})
\label{Frankeq}
\end{equation}

We use Eq.~\ref{Frankeq} to calculate $\rho$ from $2.216$ GPa up to $37.4$ GPa. The upper limit $37.4$ GPa is determined by the intersection between FFH2004's EOS and~\citep*{French:2009}'s EOS (FMNR2009). 

\subsubsection{FMNR2009's EOS}
FMNR2009's EOS is used for Ice VII, Ice X and superionic phase of H${}_{2}$O for $37.4$ GPa$\leq$p$\leq8.893$ TPa. This EOS is determined by Quantum Molecular Dynamics Simulations using the Vienna $Ab~Initio$ Simulation Package (VASP). The simulation is based on finite temperature density-functional theory (DFT) for the electronic structure and treating the ions as classical particles. Most of~\citeauthor{French:2009}'s simulations consider $54$ H${}_{2}$O molecules in a canonical ensemble, with the standard VASP PAW potentials, the $900$ eV plane-wave cutoff, and the evaluation of the electronic states at the $\Gamma$ point considered, for the $3$ independent variables: temperature (T), volume (V), and particle number (N). The simulation results are the thermal EOS p(T,V,N), and the caloric EOS U(T,V,N). The data are tabulated in FMNR2009. 

In order to approximate density $\rho$ along the melting curve, from Table V in FMNR2009, we take $1$ data point from Ice VII phase at $1000$ K and $2.5 g/cm^3$, $4$ data points from Ice X phase at $2000$ K and $3.25 g/cm^3$ to $4.00 g/cm^3$, and the rest of the data points from the superionic phase at $4000$ K and $5.00 g/cm^3$ up to $15 g/cm^3$. Since the temperature effect on density becomes smaller towards higher pressure, all the isothermal pressure-density curves converge on the isentropic pressure-density curves as well as the pressure-density curve along the melting curve.

The FMNR2009's EOS has been confirmed experimentally by Thomas Mattson et al. at the Sandia National Laboratories. At $8.893$ TPa, FMNR2009's EOS is switched to the TFD EOS in~\citealt*{Seager:2007} (SKHMM2007).

\subsubsection{SKHMM2007's EOS}
At ultrahigh pressure, the effect of electron-electron interaction can be safely ignored and electrons can be treated as a gas of non-interacting particles that obey the Pauli exclusion principle subject to the Coulomb field of the nuclei. Assuming the Coulomb potential is spatially slowly varying throughout the electron gas that the electronic wave functions can be approximated locally as plane waves, the so-called TFD solution could be derived so that the Pauli exclusion pressure balances out the Coulomb forces~\citep{Eliezer:2002,Macfarlane:1984}.

In SKHMM2007, a modified TFD by~\citet{Salpeter:1967} is used. It is modified in the sense that the authors have added in a density-dependent correlation energy term which characterizes electron interaction effects. 

Here,~\citeauthor{Seager:2007}'s EOS is slightly modified to connect to the FMNR2009's EOS. At $8.893$ TPa, the ratio of the density $\rho$ between the FMNR2009's EOS and ~\citeauthor{Seager:2007}'s EOS is $1.04464$. This ratio is multiplied to the original ~\citeauthor{Seager:2007}'s EOS density $\rho$ to produce the actual EOS used in our calculation for p$>8.893$ TPa. Only the most massive planets in our model could reach this pressure in the H${}_{2}$O-layer, so this choice of EOS for p$>8.893$ TPa has small effect on the overall mass-radius relation to be discussed in the next section. 

The EOS of H${}_{2}$O is shown in Fig.~\ref{eosplot} as the lower curve (blue curve).

\section{Result}

\subsection{Mass-Radius Contours}

Given mass and radius input, various sets of mass-radius contours can be used to quickly determine the characteristic interior structure quantities of a $2$-layer planet including its p0 (central pressure), p1/p0 (ratio of core-mantle boundary pressure over central pressure), CMF (core mass fraction), and CRF (core radius fraction). 

The $2$-layer model is uniquely solved and represented as a point on the mass-radius diagram given any pair of two parameters from the following list: $M$ (mass), $R$ (radius), p0, p1/p0, CMF, CRF, etc. The contours of constant $M$ or $R$ are trivial on the mass-radius diagram, which are simply vertical or horizontal lines. The contours of constant p0, p1/p0, CMF, or CRF are more useful. 

Within a pair of parameters, fixing one and continuously varying the other, the point on the mass-radius diagram moves to form a curve. Multiple values of the fixed parameter give multiple parallel curves forming a set of contours. The other set of contours can be obtained by exchanging the fixed parameter for the varying parameter. 
The two sets of contours crisscross each other to form a mesh, which is a natural coordinate system (see Fig.~\ref{contourplots}) of this pair of parameters, superimposed onto the existing Cartesian $(M,R)$ coordinates of the mass-radius diagram. This mesh can be used to determine the two parameters given the mass and radius input or vice versa. 

\subsubsection{Fe-MgSiO${}_{3}$ planet}

\begin{figure}[htbp]
\begin{center}
\includegraphics[scale=0.3]{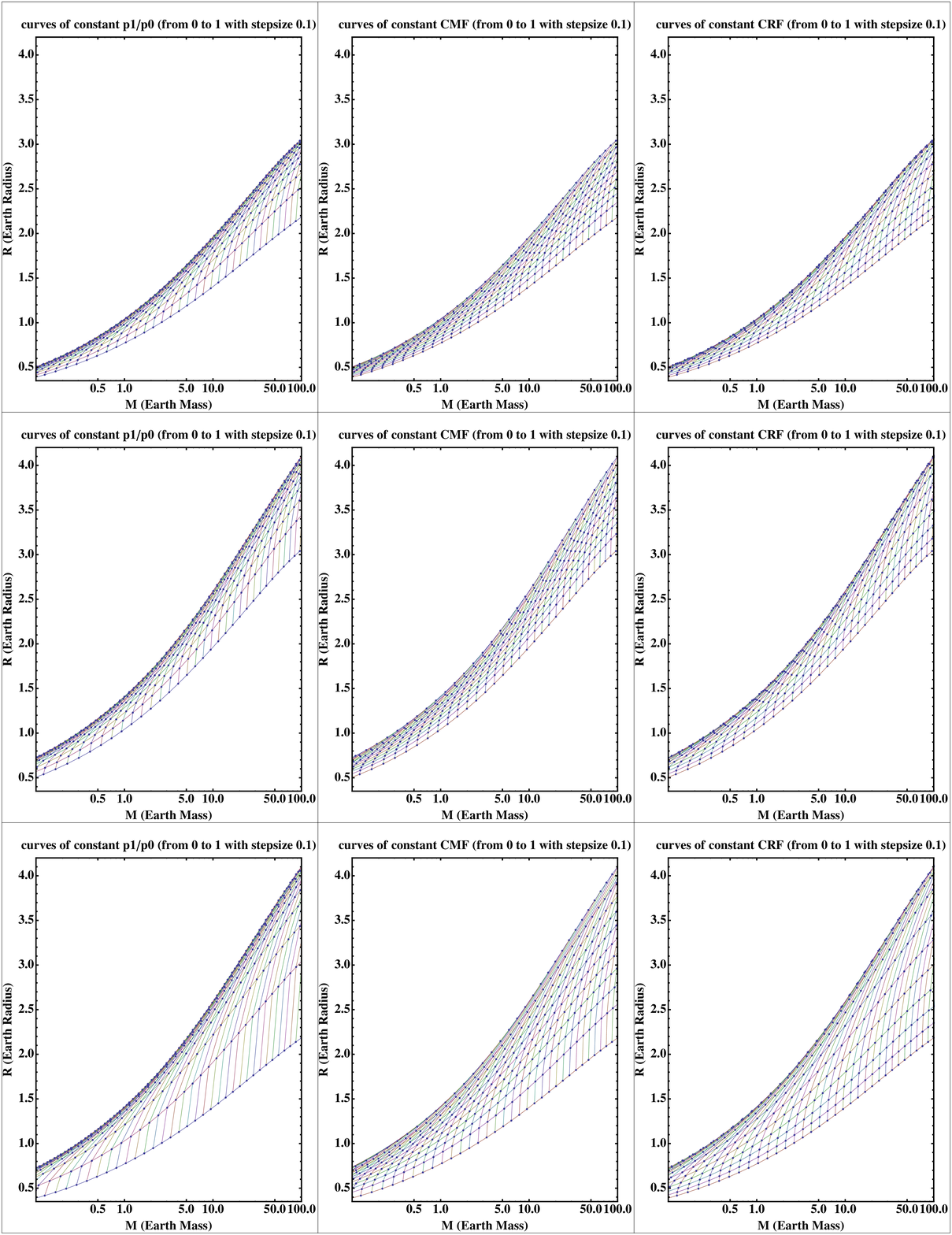}
\end{center}
\caption{Mass-Radius contours of $2$-layer planet. 1st row: Fe-MgSiO${}_{3}$ planet. 2nd row: MgSiO${}_{3}$-H${}_{2}$O planet. 3rd row: Fe-H${}_{2}$O planet. 1st column:  contour mesh of p1/p0 with p0. 2nd column:  contour mesh of CMF with p0. 3rd column: contour mesh of CRF with p0. To find out what p0 value each p0-contour corresponds to, please refer to Table~\ref{Table1}.}
\label{contourplots}
\end{figure}

As an example, the mesh of p0 with p1/p0 for Fe-core MgSiO${}_{3}$-mantle planet is illustrated in the first subplot (upper left corner) of Fig.~\ref{contourplots}. The mesh is formed by p0-contours and p1/p0-contours crisscrossing each other. The more vertical set of curves represents the p0-contours. The ratio of adjacent p0-contours is $10^{0.1}$ (see Table~\ref{Table1}). The more horizontal set of curves represents the p1/p0-contours. From bottom up, the p1/p0 values vary from 0 to 1 with step size 0.1. 

Given a pair of p0 and p1 input, users may interpolate from the mesh to find the mass and radius. On the other hand, given the mass and radius of a planet, users may also interpolate from the mesh to find the corresponding p0 and p1 of a 2-layer Fe core MgSiO${}_{3}$-mantle planet.

Similarly, the contour mesh of p0 with CMF for the Fe-MgSiO${}_{3}$ planet is shown as the second subplot from the left in the first row of Fig.~\ref{contourplots}. As a reference point, for a pure-Fe planet with $p0=10^{11}$ Pa, $M=0.1254 M_{\oplus}, R=0.417 R_{\oplus}$. 

The contour mesh of p0 with CRF for the Fe-MgSiO${}_{3}$ planet is shown as the third subplot from the left of the first row of Fig.~\ref{contourplots}.

\subsubsection{MgSiO${}_{3}$-H${}_{2}$O planet}
For $2$-layer MgSiO${}_{3}$-H${}_{2}$O planet, the $3$ diagrams (p0 contours pair with p1/p0 contours, CMF contours, or CRF contours) are the subplots of the second row of Fig.~\ref{contourplots}. 

As a reference point, for a pure-MgSiO${}_{3}$ planet with $p0=10^{10.5}$ Pa, $M=0.122 M_{\oplus}, R=0.5396 R_{\oplus}$.

\subsubsection{Fe-H${}_{2}$O planet}
For $2$-layer Fe-H${}_{2}$O planet, the $3$ diagrams (p0 contours pair with p1/p0 contours, CMF contours, or CRF contours) are the subplots of the third row of Fig.~\ref{contourplots}.

As a reference point, for a pure-Fe planet with $p0=10^{11}$ Pa, $M=0.1254 M_{\oplus}, R=0.417 R_{\oplus}$.

\subsection{Mass-Radius Curves}\label{4.2}

\begin{figure}[htbp]
\begin{center}
\includegraphics[scale=0.75]{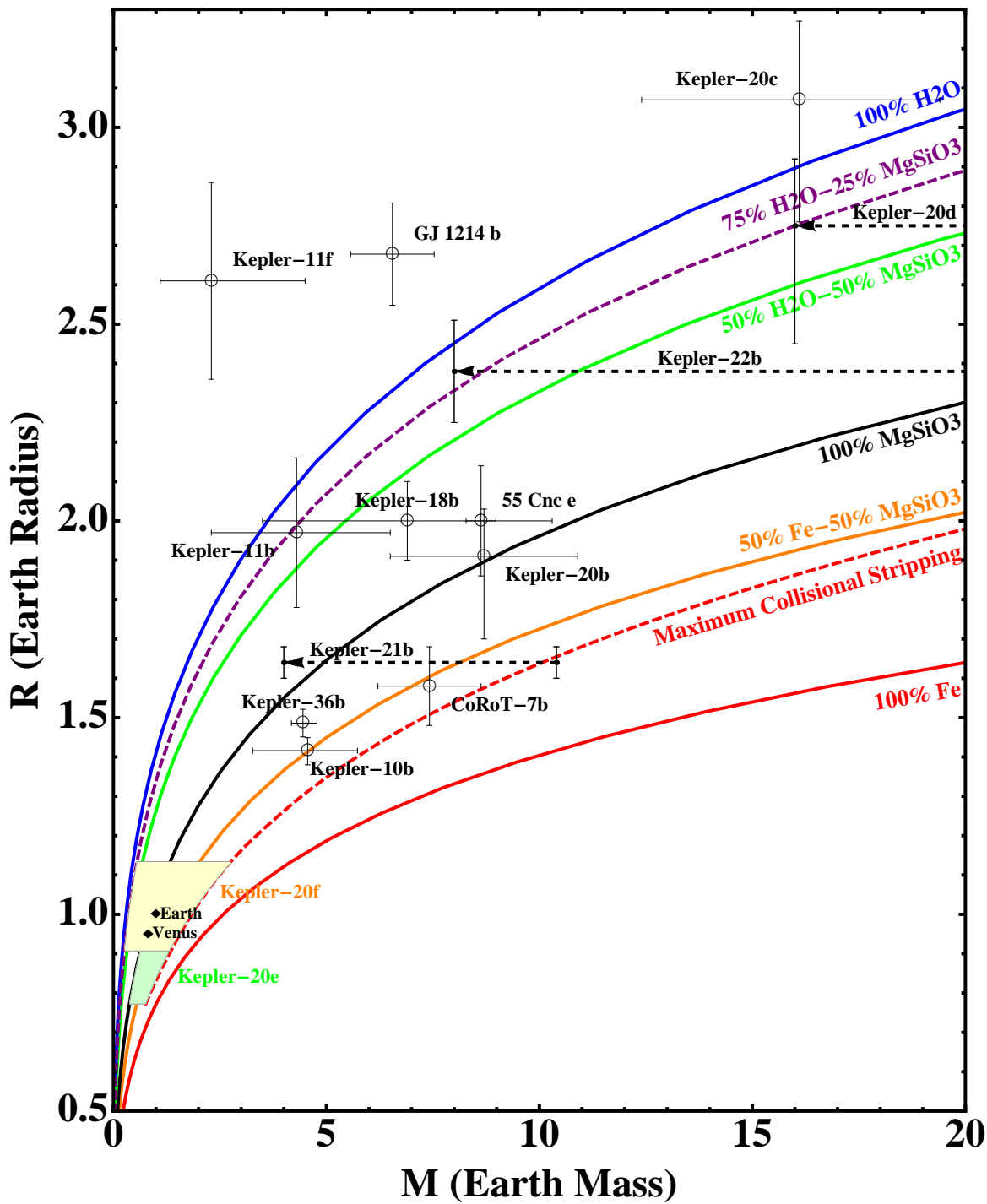}
\end{center}
\caption{Currently known transiting exoplanets are shown with their measured mass and radius with observation uncertainties. Earth and Venus are shown for comparison. The curves are calculated for planets composed of pure Fe, 50$\%$ Fe-50$\%$ MgSiO${}_{3}$, pure MgSiO${}_{3}$, 50$\%$ H${}_{2}$O-50$\%$ MgSiO${}_{3}$, 75$\%$ H${}_{2}$O-25$\%$ MgSiO${}_{3}$ and pure H${}_{2}$O. The red dashed curve is the maximum collisional stripping curve calculated by~\citet{Marcus:2010}.} 
\label{MRplot}
\end{figure}

For observers' interest, $6$ characteristic mass-radius curves are plotted (Fig.~\ref{MRplot}) and tabulated (Table~\ref{Table1}), representing the pure-Fe planet, half-Fe half-MgSiO${}_{3}$ planet, pure MgSiO${}_{3}$ planet, half-MgSiO${}_{3}$ half-H${}_{2}$O planet, 75$\%$ H${}_{2}$O-25$\%$ MgSiO${}_{3}$ planet, and pure H${}_{2}$O planet. These fractions are mass fractions. Fig.~\ref{MRplot} also shows some recently discovered exoplanets within the relevant mass-radius regime for comparison. These planets include Kepler-10b~\citep[][]{Batalha:2011}, Kepler-11b~\citep[][]{Lissauer:2011}, Kepler-11f~\citep[][]{Lissauer:2011}, Kepler-18b~\citep[][]{Cochran:2011}, Kepler-36b~\citep[][]{Carter:2012}, and Kepler-20b,c,d~\citep[][]{Gautier:2012}. They also include Kepler-20e ($R=0.868_{-0.096}^{+0.074} R_{\oplus}$~\citep[][]{Fressin:2011}, the mass range is determined by pure-silicate mass-radius curve and the maximum collisional stripping curve~\citep[][]{Marcus:2010}), Kepler-20f ($R=1.034_{-0.127}^{+0.100} R_{\oplus}$~\citep[][]{Fressin:2011}, the mass range is determined by $75\%$ water-ice and $25\%$ silicate mass-radius curves and the maximum collisional stripping curve~\citep[][]{Marcus:2010}), Kepler-21b ($R=1.64\pm0.04 R_{\oplus}$~\citep[][]{Howell:2012}, The upper limit for mass is $10.4 M_{\oplus}$: the 2-$\sigma$ upper limit preferred in the paper. The lower limit is $4 M_{\oplus}$, which is in between the "Earth" and "50$\%$ H${}_{2}$O-50$\%$ MgSiO${}_{3}$" model curves - the planet is very hot and is unlikely to have much water content if any at all.), Kepler-22b ($R=2.38\pm0.13 R_{\oplus}$~\citep[][]{Borucki:2012}, The 1-$\sigma$ upper limit for mass is $36 M_{\oplus}$ for an eccentric orbit (or $27 M_{\oplus}$, for circular orbit)), CoRoT-7b ($M=7.42\pm1.21M_{\oplus}$, $R=1.58\pm0.1 R_{\oplus}$~\citep[][]{Hatzes:2011, Leger:2009, Queloz:2009}), 55 Cancri e ($M=8.63\pm0.35M_{\oplus}$, $R=2.00\pm0.14 R_{\oplus}$~\citep[][]{Winn:2011}), and GJ 1214b~\citep[][]{Charbonneau:2009}. 

\begin{sidewaystable}[htbp]
\caption{Data for the 6 characteristic mass-radius curves\label{Table1}}

\centering
\scalebox{0.65}{
\begin{tabular}{c | c c | c c c c | c c | c c c c | c c c c | c c |}
\\
\hline\hline
  &  
 \multicolumn{2}{c}{Fe} & 
 \multicolumn{4}{c}{$50\% Fe$-$50\% MgSiO_3$} & 
 \multicolumn{2}{c}{MgSiO${}_{3}$} & 
 \multicolumn{4}{c}{$50\% MgSiO_3$-$50\% H_2O$} & 
 \multicolumn{4}{c}{$25\% MgSiO_3$-$75\% H_2O$} &
 \multicolumn{2}{c}{H${}_{2}$O} \\ \hline
 
$log_{10}(p0)$ & Mass\footnotemark[1] & Radius\footnotemark[2] & Mass & Radius & CRF\footnotemark[3] & p1/p0\footnotemark[4] & Mass & Radius & Mass & Radius & CRF & p1/p0 & Mass & Radius & CRF & p1/p0 & Mass & Radius\\ \hline

9.6	&	0.001496	&	0.09947	&	0.00177	&	0.121	&	0.6892	&	0.2931	&	0.00623	&	0.2029	&	0.008278	&	0.2963	&	0.5969	&	0.2374	&	0.01217	&	0.3616	&	0.4413	&	0.3782	&	0.04663	&	0.58	\\
9.7	&	0.002096	&	0.1112	&	0.002481	&	0.1354	&	0.6888	&	0.2925	&	0.008748	&	0.227	&	0.01156	&	0.3286	&	0.6013	&	0.2398	&	0.01699	&	0.4009	&	0.4444	&	0.3803	&	0.06174	&	0.63	\\
9.8	&	0.002931	&	0.1243	&	0.003472	&	0.1513	&	0.6882	&	0.2918	&	0.01227	&	0.254	&	0.01615	&	0.3647	&	0.6051	&	0.2413	&	0.02351	&	0.4432	&	0.4475	&	0.385	&	0.08208	&	0.6853	\\
9.9	&	0.00409	&	0.1387	&	0.004848	&	0.169	&	0.6876	&	0.2909	&	0.01717	&	0.2838	&	0.02255	&	0.4049	&	0.6085	&	0.242	&	0.03213	&	0.4869	&	0.4513	&	0.3944	&	0.1091	&	0.7455	\\
10	&	0.005694	&	0.1546	&	0.006752	&	0.1885	&	0.6868	&	0.2898	&	0.02399	&	0.317	&	0.0313	&	0.4484	&	0.6119	&	0.2446	&	0.04399	&	0.535	&	0.4554	&	0.4016	&	0.1445	&	0.8102	\\
10.1	&	0.007904	&	0.1722	&	0.00938	&	0.2101	&	0.6858	&	0.2886	&	0.03343	&	0.3536	&	0.04314	&	0.4944	&	0.6165	&	0.2495	&	0.06029	&	0.588	&	0.4592	&	0.4073	&	0.1904	&	0.879	\\
10.2	&	0.01094	&	0.1915	&	0.01299	&	0.2339	&	0.6847	&	0.2871	&	0.04644	&	0.3939	&	0.05943	&	0.5449	&	0.6209	&	0.253	&	0.08255	&	0.6461	&	0.4629	&	0.4119	&	0.2494	&	0.9515	\\
10.3	&	0.01507	&	0.2126	&	0.01792	&	0.26	&	0.6834	&	0.2855	&	0.0643	&	0.4381	&	0.0817	&	0.6003	&	0.6249	&	0.2554	&	0.1127	&	0.7093	&	0.4662	&	0.4159	&	0.3249	&	1.028	\\
10.4	&	0.0207	&	0.2356	&	0.02464	&	0.2886	&	0.6819	&	0.2836	&	0.08866	&	0.4865	&	0.112	&	0.6607	&	0.6285	&	0.2571	&	0.1533	&	0.7777	&	0.4693	&	0.4197	&	0.4206	&	1.107	\\
10.5	&	0.02829	&	0.2606	&	0.03372	&	0.3197	&	0.6801	&	0.2815	&	0.1217	&	0.5391	&	0.1528	&	0.7262	&	0.6318	&	0.2585	&	0.2074	&	0.8509	&	0.4721	&	0.4234	&	0.5416	&	1.19	\\
10.6	&	0.0385	&	0.2876	&	0.04595	&	0.3535	&	0.6782	&	0.2792	&	0.1661	&	0.596	&	0.2074	&	0.7966	&	0.6347	&	0.2595	&	0.2789	&	0.9289	&	0.4747	&	0.427	&	0.6938	&	1.276	\\
10.7	&	0.05212	&	0.3167	&	0.06231	&	0.3901	&	0.6759	&	0.2767	&	0.2255	&	0.6573	&	0.2799	&	0.8718	&	0.6373	&	0.2603	&	0.3726	&	1.011	&	0.477	&	0.4305	&	0.8866	&	1.366	\\
10.8	&	0.07021	&	0.348	&	0.08408	&	0.4296	&	0.6735	&	0.274	&	0.3042	&	0.7228	&	0.3754	&	0.9515	&	0.6396	&	0.261	&	0.4946	&	1.098	&	0.4792	&	0.4339	&	1.132	&	1.461	\\
10.9	&	0.09408	&	0.3814	&	0.1129	&	0.4721	&	0.6708	&	0.2711	&	0.4075	&	0.7925	&	0.4998	&	1.035	&	0.6416	&	0.2615	&	0.6517	&	1.188	&	0.4811	&	0.437	&	1.444	&	1.562	\\
11	&	0.1254	&	0.417	&	0.1508	&	0.5177	&	0.6679	&	0.2682	&	0.542	&	0.8661	&	0.6607	&	1.123	&	0.6434	&	0.2618	&	0.8529	&	1.282	&	0.4827	&	0.4397	&	1.841	&	1.669	\\
11.1	&	0.1663	&	0.4548	&	0.2003	&	0.5663	&	0.6648	&	0.2651	&	0.7143	&	0.9429	&	0.8653	&	1.214	&	0.6448	&	0.2615	&	1.107	&	1.38	&	0.4839	&	0.4411	&	2.346	&	1.782	\\
11.2	&	0.2193	&	0.4949	&	0.2647	&	0.618	&	0.6616	&	0.2621	&	0.927	&	1.02	&	1.117	&	1.305	&	0.6455	&	0.2598	&	1.42	&	1.477	&	0.4843	&	0.44	&	2.985	&	1.901	\\
11.3	&	0.2877	&	0.537	&	0.348	&	0.6728	&	0.6582	&	0.259	&	1.2	&	1.101	&	1.437	&	1.4	&	0.6458	&	0.2589	&	1.818	&	1.58	&	0.484	&	0.44	&	3.77	&	2.023	\\
11.4	&	0.3754	&	0.5814	&	0.455	&	0.7307	&	0.6547	&	0.2559	&	1.545	&	1.186	&	1.842	&	1.499	&	0.6458	&	0.2582	&	2.321	&	1.688	&	0.4834	&	0.4403	&	4.735	&	2.147	\\
11.5	&	0.4875	&	0.6279	&	0.5919	&	0.7916	&	0.6512	&	0.2529	&	1.981	&	1.274	&	2.351	&	1.602	&	0.6453	&	0.2574	&	2.957	&	1.802	&	0.4828	&	0.4399	&	5.909	&	2.274	\\
11.6	&	0.6298	&	0.6765	&	0.7659	&	0.8554	&	0.6476	&	0.25	&	2.525	&	1.365	&	2.988	&	1.708	&	0.6445	&	0.2565	&	3.752	&	1.92	&	0.4822	&	0.4391	&	7.325	&	2.401	\\
11.7	&	0.8096	&	0.727	&	0.986	&	0.9221	&	0.644	&	0.2473	&	3.203	&	1.458	&	3.78	&	1.818	&	0.6435	&	0.2552	&	4.732	&	2.041	&	0.4813	&	0.4383	&	9.038	&	2.529	\\
11.8	&	1.036	&	0.7796	&	1.261	&	0.9907	&	0.6407	&	0.2451	&	4.043	&	1.554	&	4.763	&	1.933	&	0.6423	&	0.2536	&	5.936	&	2.164	&	0.4803	&	0.4377	&	11.11	&	2.66	\\
11.9	&	1.319	&	0.834	&	1.606	&	1.062	&	0.6374	&	0.2429	&	5.077	&	1.653	&	5.972	&	2.05	&	0.641	&	0.252	&	7.407	&	2.289	&	0.4792	&	0.4373	&	13.55	&	2.789	\\
12	&	1.671	&	0.8902	&	2.036	&	1.136	&	0.6342	&	0.2407	&	6.297	&	1.749	&	7.392	&	2.165	&	0.6394	&	0.2488	&	9.127	&	2.411	&	0.4778	&	0.4341	&	16.42	&	2.915	\\
12.1	&	2.108	&	0.9481	&	2.568	&	1.211	&	0.631	&	0.2385	&	7.714	&	1.842	&	9.043	&	2.275	&	0.6372	&	0.2449	&	11.12	&	2.529	&	0.4755	&	0.4299	&	19.77	&	3.039	\\
12.2	&	2.648	&	1.007	&	3.226	&	1.289	&	0.6278	&	0.2363	&	9.423	&	1.935	&	11.03	&	2.386	&	0.6345	&	0.242	&	13.52	&	2.647	&	0.4727	&	0.4275	&	23.68	&	3.16	\\
12.3	&	3.31	&	1.068	&	4.032	&	1.369	&	0.6247	&	0.2342	&	11.47	&	2.029	&	13.4	&	2.498	&	0.6317	&	0.2396	&	16.37	&	2.765	&	0.4695	&	0.4265	&	28.21	&	3.278	\\
12.4	&	4.119	&	1.13	&	5.018	&	1.451	&	0.6216	&	0.2321	&	13.87	&	2.121	&	16.18	&	2.608	&	0.6285	&	0.2371	&	19.72	&	2.882	&	0.467	&	0.4248	&	33.49	&	3.393	\\
12.5	&	5.103	&	1.193	&	6.216	&	1.534	&	0.6184	&	0.23	&	16.73	&	2.213	&	19.48	&	2.717	&	0.6252	&	0.2353	&	23.68	&	2.998	&	0.4646	&	0.4237	&	39.62	&	3.506	\\
12.6	&	6.293	&	1.257	&	7.665	&	1.618	&	0.6153	&	0.228	&	20.1	&	2.304	&	23.36	&	2.825	&	0.6219	&	0.2339	&	28.31	&	3.111	&	0.4623	&	0.423	&	46.72	&	3.616	\\
12.7	&	7.727	&	1.321	&	9.39	&	1.7	&	0.6126	&	0.2267	&	24.07	&	2.394	&	27.94	&	2.933	&	0.619	&	0.2323	&	33.71	&	3.222	&	0.46	&	0.4225	&	54.92	&	3.724	\\
12.8	&	9.445	&	1.386	&	11.45	&	1.783	&	0.61	&	0.2254	&	28.68	&	2.483	&	33.24	&	3.037	&	0.6162	&	0.2306	&	39.97	&	3.33	&	0.4579	&	0.4214	&	64.22	&	3.826	\\
12.9	&	11.49	&	1.451	&	13.91	&	1.865	&	0.6074	&	0.2241	&	33.99	&	2.569	&	39.33	&	3.138	&	0.6134	&	0.2288	&	47.15	&	3.434	&	0.4556	&	0.4199	&	74.79	&	3.924	\\
13	&	13.92	&	1.515	&	16.81	&	1.947	&	0.6048	&	0.2227	&	40.05	&	2.65	&	46.26	&	3.234	&	0.6105	&	0.2266	&	55.31	&	3.534	&	0.4531	&	0.4177	&	86.85	&	4.017	\\
13.1	&	16.78	&	1.579	&	20.21	&	2.027	&	0.6024	&	0.2213	&	46.88	&	2.728	&	54.07	&	3.325	&	0.6075	&	0.2238	&	64.47	&	3.627	&	0.4503	&	0.4148	&	100.3	&	4.104	\\
13.2	&	20.14	&	1.642	&	24.2	&	2.106	&	0.5999	&	0.2198	&	54.49	&	2.799	&	62.77	&	3.409	&	0.6042	&	0.2207	&	74.65	&	3.713	&	0.4472	&	0.4114	&	115.3	&	4.183	\\
13.3	&	24.05	&	1.704	&	28.85	&	2.183	&	0.5973	&	0.2182	&	63.08	&	2.866	&	72.58	&	3.488	&	0.6006	&	0.2178	&	86.14	&	3.793	&	0.4438	&	0.4087	&	131.9	&	4.256	\\
13.4	&	28.6	&	1.765	&	34.23	&	2.258	&	0.5947	&	0.2166	&	72.75	&	2.928	&	83.62	&	3.563	&	0.5967	&	0.2152	&	99.08	&	3.87	&	0.4404	&	0.4064	&	150.3	&	4.322	\\
13.5	&	33.87	&	1.824	&	40.45	&	2.331	&	0.5921	&	0.2151	&	83.59	&	2.986	&	95.97	&	3.631	&	0.5928	&	0.2129	&	113.6	&	3.94	&	0.437	&	0.4043	&	170.8	&	4.382	\\
13.6	&	39.94	&	1.881	&	47.59	&	2.401	&	0.5895	&	0.2136	&	95.68	&	3.039	&	109.8	&	3.695	&	0.5888	&	0.2106	&	129.7	&	4.006	&	0.4338	&	0.4022	&	193.6	&	4.435	\\
13.7	&	46.92	&	1.937	&	55.76	&	2.468	&	0.5869	&	0.2123	&	109.1	&	3.088	&	125.1	&	3.754	&	0.5847	&	0.2083	&	147.6	&	4.064	&	0.4306	&	0.4001	&	218.7	&	4.483	\\
13.8	&	54.93	&	1.99	&	65.07	&	2.531	&	0.5844	&	0.2114	&	124	&	3.132	&	142	&	3.807	&	0.5807	&	0.2061	&	167.2	&	4.116	&	0.4274	&	0.3979	&	246.6	&	4.525	\\
13.9	&	64.08	&	2.042	&	75.65	&	2.59	&	0.582	&	0.2106	&	140.3	&	3.17	&	160.6	&	3.854	&	0.5768	&	0.2038	&	188.8	&	4.162	&	0.4242	&	0.3956	&	277.3	&	4.561	\\
14	&	74.51	&	2.091	&	87.65	&	2.645	&	0.5797	&	0.21	&	158.2	&	3.204	&	181	&	3.895	&	0.5728	&	0.2015	&	212.5	&	4.202	&	0.4209	&	0.3932	&	311.3	&	4.592	\\
14.1	&	86.37	&	2.138	&	101.2	&	2.697	&	0.5775	&	0.2094	&	177.8	&	3.232	&	203.3	&	3.93	&	0.5688	&	0.1992	&	238.4	&	4.236	&	0.4175	&	0.3908	&	348.7	&	4.618	\\
14.2	&	99.8	&	2.183	&	116.5	&	2.745	&	0.5755	&	0.2089	&	199.2	&	3.255	&	227.7	&	3.959	&	0.5648	&	0.1971	&	266.8	&	4.265	&	0.4142	&	0.3884	&	390.1	&	4.639	\\
14.3	&	115	&	2.226	&	133.8	&	2.789	&	0.5735	&	0.2085	&	222.5	&	3.274	&	254.2	&	3.983	&	0.5607	&	0.1951	&	297.9	&	4.288	&	0.4109	&	0.3861	&	435.9	&	4.656	\\
14.4	&	132.1	&	2.266	&	153.1	&	2.829	&	0.5716	&	0.2082	&	248.1	&	3.288	&	283.3	&	4.002	&	0.5567	&	0.1932	&	332.1	&	4.307	&	0.4076	&	0.3841	&	486.4	&	4.669	\\

\hline
\end{tabular}
}

\end{sidewaystable}

\footnotetext[1]{mass in Earth Mass ($M_{\oplus} = 5.9742\times10^{24} kg$)}
\footnotetext[2]{radius in Earth Radius ($R_{\oplus} = 6.371\times10^{6} m$)}
\footnotetext[3]{CRF stands for core radius fraction, the ratio of the radius of the core over the total radius of the planet, in the two-layer model}
\footnotetext[4]{p1/p0 stands for core-mantle-boundary pressure fraction, the ratio of the pressure at the core-mantle boundary (p1) over the pressure at the center of the planet (p0), in the two-layer model}

\subsection{Levels of planet differentiation: the effect of Fe partitioning between mantle and core\label{nondiff}}

All models of planets discussed so far assume that all Fe is in the core, while all Mg, Si and O are in the mantle, i.e., that a planet is fully differentiated. However, we know that in terrestrial planets some Fe is incorporated into the mantle. There are two separate processes which affect the Fe content of the mantle: (1) mechanical segregation of Fe-rich metal from the mantle to the core, and (2) different redox conditions resulting in a different Fe/Mg ratio within the mantle, which in turn affects the relative size of the core and mantle. In this section we show the effects of (1) undifferentiated versus fully-differentiated, and (2) reduced versus oxidized planetary structure on the mass-radius relation for a planet with the same Fe/Si and Mg/Si ratios. 

For simplicity, here we ignore the H${}_{2}$O and gaseous content of the planet and only consider the planet made of Fe, Mg, Si, O. To facilitate comparison between different cases, we fix the global atomic ratios of Fe/Mg=1 and Mg/Si=1; these fit well within the range of local stellar abundances~\citep{Grasset:2009}. 

In particular, we consider a double-layer planet with a core and a mantle in two scenarios. 
One follows the incomplete mechanical separation of the Fe-rich metal during planet formation, and results in addition of Fe to the mantle as metal particles. It does not change EOS of the silicate components, but requires adding an Fe-EOS to the mantle mixture. Thus, the planet generally consists of a Fe metal core and a partially differentiated mantle consisting of the mixture of metallic Fe and MgSiO${}_{3}$ silicates. While the distribution of metallic Fe may have a radial gradient, for simplicity we assume that it is uniformly distributed in the silicate mantle. Within scenario 1, we calculate three cases to represent different levels of differentiation: 

\begin{description}
\item[Case 1: complete differentiation] metallic Fe core and MgSiO${}_{3}$ silicate mantle. For Fe/Mg=1, CMF=0.3574. 
\item[Case 2: partial differentiation] half the Fe forms a smaller metallic Fe core, with the other half of metal being mixed with MgSiO${}_{3}$ silicates in the mantle. CMF=0.1787. 
\item[Case 3: no differentiation] All the metallic Fe is mixed with MgSiO${}_{3}$ in the mantle. CMF=0 (no core). 
\end{description}

The other scenario assumes different redox conditions, resulting in different Fe/Mg ratios in mantle minerals, and therefore requiring different EOS for the Fe${}^{2+}$-bearing silicates and oxides. More oxidized mantle means adding more Fe in the form of FeO to the MgSiO${}_{3}$ silicates to form (Mg,Fe)SiO${}_{3}$ silicates and (Mg,Fe)O magnesiow\"{u}stite (mv), thus reducing the amount of Fe in the core. The exact amounts of (Mg,Fe)SiO${}_{3}$ and (Mg,Fe)O in the mantle are determined by the following mass balance equation: 

\begin{equation}
x~FeO + MgSiO3~\xrightarrow~(Mg{}_{\frac{1}{1+x}},Fe{}_{\frac{x}{1+x}})SiO{}_3 + x~(Mg_{\frac{1}{1+x}},Fe_{\frac{x}{1+x}})O
\label{massbalance}
\end{equation}

$x$ denotes the relative amount of FeO added to the silicate mantle, $x$=0 being the most reduced state with no Fe in the mantle, and $x$=1 being the most oxidized state with all Fe existing as oxides in the mantle. This oxidization process conserves the global Fe/Mg and Mg/Si ratios, but increases O content and thus the O/Si ratio of the planet since Fe is added to the mantle in the form of FeO. Because in stellar environments O is excessively abundant relative to Mg, Si, and Fe (e.g., the solar elemental abundances~\citep{Asplund:2009}), it is not a limiting factor in our models of oxidized planets. 
We calculate the following three cases to represent the full range of redox conditions: 

\begin{description}
\item[Case 4: no oxidization of Fe] $x$=0. Metal Fe core and MgSiO${}_{3}$ silicate mantle. For Fe/Mg=1, O/Si=3, CMF=0.3574. 
\item[Case 5: partial oxidization of Fe] $x$=0.5. Half the Fe forms smaller metal core, the other half is added as FeO to the mantle. O/Si=3.5, CMF=0.1700. 
\item[Case 6: complete oxidization of Fe] $x$=1. All Fe is added as FeO to the mantle, resulting in no metal core at all. O/Si=4, CMF=0. 
\end{description}

Notice that Case 4 looks identical to Case 1. However, the silicate EOS used to calculate Case 4 is different from Case 1 at ultrahigh pressures (beyond $0.90$ TPa). For Cases 4, 5, and 6, the (Mg,Fe)SiO${}_{3}$ EOS is adopted from~\citet{Caracas:2008} and~\citet{Spera:2006} which only consider perovskite (pv) and post-perovskite (ppv) phases without including further dissociation beyond $0.90$ TPa, since the Fe${}^{2+}$-bearing silicate EOS at ultrahigh pressures is hardly available. On the other hand, Comparison between Case 1 and Case 4 also shows the uncertainty on mass-radius relation resulting from the different choice of EOS (see Table~\ref{Table2}). Fe${}^{2+}$-bearing pv and ppv have the general formula: (Mg${}_{y}$,Fe${}_{1-y}$)SiO${}_{3}$, where $y$ denotes the relative atomic number fraction of Mg and Fe in the silicate mineral. The (Mg${}_{y}$,Fe${}_{1-y}$)SiO${}_{3}$ silicate is therefore a binary component equilibrium solid solution. It could either be pv or ppv or both depending on the pressure~\citep{Spera:2006}. We can safely approximate the narrow pressure region where pv and ppv co-exist as a single transition pressure from pv to ppv. This pressure is calculated as the arithmetic mean of the initial transition pressure of pv $\rightarrow$ pv+ppv mixture and the final transition pressure of pv+ppv mixture $\rightarrow$ ppv~\citep{Spera:2006}. The pv EOS and ppv EOS are connected at this transition pressure to form a consistent EOS for all pressures. 

Addition of FeO to MgSiO${}_{3}$ results in the appearance of a second phase, magnesiow\"{u}stite, in the mantle according to Eq.~\ref{massbalance}. The (Mg,Fe)O EOS for Cases 4, 5, and 6 is adopted from~\citet{Fei:2007}, which includes the electronic spin transition of high-spin to low-spin in Fe${}^{2+}$. For simplicity, we assume that pv/ppv and mw have the same Mg/Fe ratio. 

\begin{figure}[htbp]
\begin{center}
\includegraphics[scale=0.75]{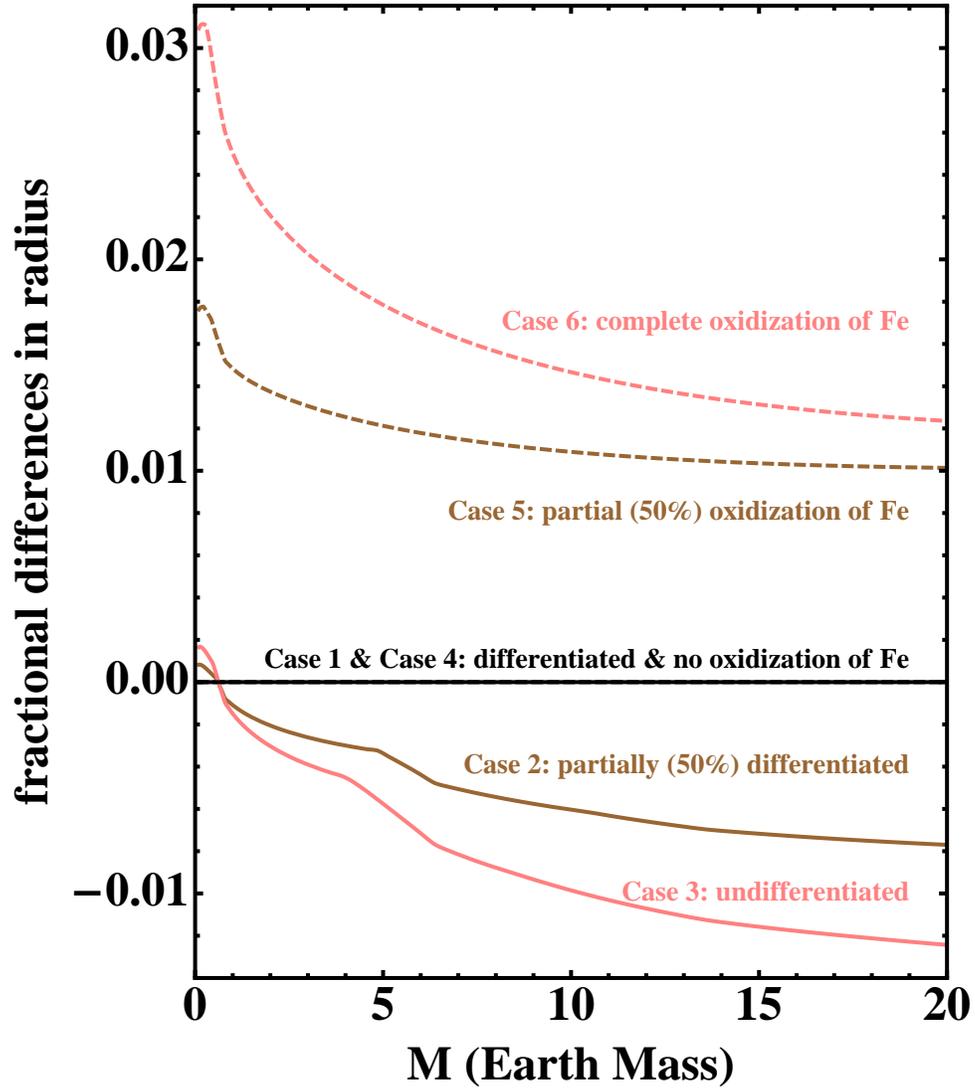}
\end{center}
\caption{fractional differences in radius resulting from Fe partitioning between mantle and core. 
black curve: Case 1 \& Case 4 (complete differentiation and no oxidization of Fe); solid brown curve: Case 2 (partial (50\%) differentiation: 50$\%$ metallic Fe mixed with the mantle); solid pink curve: Case 3 (no differentiation: all metallic Fe mixed with the mantle); dashed brown curve: Case 5 (partial (50\%) oxidization of Fe); dashed pink curve: Case 6 (complete oxidization of Fe)} 
\label{f4}
\end{figure}

Fig.~\ref{f4} shows fractional differences ($\eta$) in radius of Cases 2 \& 3 compared to Case 1 as well as Cases 5 \& 6 compared to Case 4 ($r_{0}$ is radius of the reference case, which is that of Case 1 for Cases 2 \& 3 and is that of Case 4 for Case 5 \& 6):
\begin{equation}
\eta=\frac{r-r_{0}}{r_{0}}
\label{fractionaldiff}
\end{equation}

Oxidization of Fe (partitioning Fe as Fe-oxides from the core into the mantle) makes the planet appear larger. The complete oxidization of Fe makes the radius 3$\%$ larger for small planets around 1 $M_{\oplus}$, then the difference decreases with increasing mass within the mass range of 1 to 20 $M_{\oplus}$. Undifferentiated planets (partitioning of metallic Fe from the core into the mantle) appear smaller than fully differentiated planets. The completely undifferentiated planet is practically indistinguishable in radius for small planets around 1 $M_{\oplus}$, then the difference increases to 1$\%$-level around 20 $M_{\oplus}$. The mass, radius, CRF and p1/p0 data of Cases 1 through 6 are listed in Table~\ref{Table2}. 

\begin{sidewaystable}[htbp]
\caption{Data of Cases 1 through 6\label{Table2}}

\centering
\scalebox{0.60}{
\begin{tabular}{c | c c c c | c c c c | c c | c c c c | c c c c | c c |}
\\
\hline\hline
  &  
 \multicolumn{4}{c}{Case 1 (CMF=0.3574)} & 
 \multicolumn{4}{c}{Case 2 (CMF=0.1787)} & 
 \multicolumn{2}{c}{Case 3} & 
 \multicolumn{4}{c}{Case 4 (CMF=0.3574)} & 
 \multicolumn{4}{c}{Case 5 (CMF=0.1700)} &
 \multicolumn{2}{c}{Case 6} \\ \hline
 
$log_{10}(p0)$ & Mass\footnotemark[1] & Radius\footnotemark[2] & CRF\footnotemark[3] & p1/p0\footnotemark[4] & Mass & Radius & CRF & p1/p0 & Mass & Radius & Mass & Radius & CRF & p1/p0 & Mass & Radius & CRF & p1/p0 & Mass & Radius\\ \hline

9.6	&	0.002009	&	0.1302	&	0.5972	&	0.3847	&	0.002395	&	0.1381	&	0.4736	&	0.5634	&	0.004165	&	0.1659	&	0.002009	&	0.1302	&	0.5972	&	0.3847	&	0.002526	&	0.1428	&	0.4584	&	0.5625	&	0.004921	&	0.1804	\\
9.7	&	0.002816	&	0.1456	&	0.5967	&	0.3839	&	0.003359	&	0.1545	&	0.4731	&	0.5626	&	0.005846	&	0.1856	&	0.002816	&	0.1456	&	0.5967	&	0.3839	&	0.003542	&	0.1597	&	0.4579	&	0.5617	&	0.006905	&	0.2018	\\
9.8	&	0.003941	&	0.1628	&	0.5961	&	0.383	&	0.004704	&	0.1727	&	0.4726	&	0.5616	&	0.008194	&	0.2076	&	0.003941	&	0.1628	&	0.5961	&	0.383	&	0.004959	&	0.1786	&	0.4574	&	0.5607	&	0.009673	&	0.2256	\\
9.9	&	0.005504	&	0.1819	&	0.5954	&	0.382	&	0.006573	&	0.1929	&	0.4719	&	0.5604	&	0.01146	&	0.232	&	0.005504	&	0.1819	&	0.5954	&	0.382	&	0.00693	&	0.1995	&	0.4567	&	0.5596	&	0.01352	&	0.252	\\
10	&	0.00767	&	0.203	&	0.5946	&	0.3807	&	0.009165	&	0.2154	&	0.4711	&	0.5591	&	0.016	&	0.2589	&	0.00767	&	0.203	&	0.5946	&	0.3807	&	0.009662	&	0.2227	&	0.456	&	0.5582	&	0.01887	&	0.2812	\\
10.1	&	0.01066	&	0.2263	&	0.5936	&	0.3792	&	0.01275	&	0.2401	&	0.4702	&	0.5575	&	0.02228	&	0.2887	&	0.01066	&	0.2263	&	0.5936	&	0.3792	&	0.01344	&	0.2484	&	0.455	&	0.5567	&	0.02625	&	0.3135	\\
10.2	&	0.01477	&	0.252	&	0.5924	&	0.3774	&	0.01767	&	0.2674	&	0.469	&	0.5557	&	0.03093	&	0.3215	&	0.01477	&	0.252	&	0.5924	&	0.3774	&	0.01863	&	0.2766	&	0.454	&	0.5548	&	0.03639	&	0.3489	\\
10.3	&	0.02039	&	0.2802	&	0.591	&	0.3754	&	0.02442	&	0.2974	&	0.4678	&	0.5536	&	0.04278	&	0.3575	&	0.02039	&	0.2802	&	0.591	&	0.3754	&	0.02574	&	0.3076	&	0.4527	&	0.5528	&	0.05027	&	0.3878	\\
10.4	&	0.02805	&	0.311	&	0.5894	&	0.3732	&	0.03362	&	0.3303	&	0.4663	&	0.5513	&	0.05893	&	0.3968	&	0.02805	&	0.311	&	0.5894	&	0.3732	&	0.03544	&	0.3416	&	0.4513	&	0.5504	&	0.06915	&	0.4301	\\
10.5	&	0.03842	&	0.3447	&	0.5876	&	0.3706	&	0.0461	&	0.3661	&	0.4647	&	0.5487	&	0.08081	&	0.4395	&	0.03842	&	0.3447	&	0.5876	&	0.3706	&	0.04859	&	0.3786	&	0.4497	&	0.5479	&	0.09465	&	0.476	\\
10.6	&	0.05239	&	0.3814	&	0.5856	&	0.3679	&	0.06292	&	0.405	&	0.4628	&	0.5459	&	0.1102	&	0.4858	&	0.05239	&	0.3814	&	0.5856	&	0.3679	&	0.06631	&	0.4189	&	0.448	&	0.5451	&	0.1289	&	0.5257	\\
10.7	&	0.07111	&	0.4211	&	0.5833	&	0.3649	&	0.08548	&	0.4472	&	0.4608	&	0.5429	&	0.1496	&	0.5356	&	0.07111	&	0.4211	&	0.5833	&	0.3649	&	0.09006	&	0.4624	&	0.446	&	0.5422	&	0.1744	&	0.5789	\\
10.8	&	0.09605	&	0.464	&	0.5809	&	0.3617	&	0.1155	&	0.4927	&	0.4587	&	0.5398	&	0.2017	&	0.5888	&	0.09605	&	0.464	&	0.5809	&	0.3617	&	0.1217	&	0.5094	&	0.4439	&	0.5392	&	0.2345	&	0.6358	\\
10.9	&	0.1291	&	0.5101	&	0.5782	&	0.3584	&	0.1553	&	0.5416	&	0.4564	&	0.5366	&	0.2702	&	0.6456	&	0.1291	&	0.5101	&	0.5782	&	0.3584	&	0.1636	&	0.5599	&	0.4417	&	0.5361	&	0.3113	&	0.6946	\\
11	&	0.1725	&	0.5595	&	0.5753	&	0.355	&	0.2077	&	0.5939	&	0.454	&	0.5334	&	0.3596	&	0.7056	&	0.1725	&	0.5595	&	0.5753	&	0.355	&	0.2187	&	0.6139	&	0.4394	&	0.533	&	0.408	&	0.7548	\\
11.1	&	0.2294	&	0.6123	&	0.5723	&	0.3515	&	0.2762	&	0.6495	&	0.4515	&	0.5302	&	0.4744	&	0.7684	&	0.2294	&	0.6123	&	0.5723	&	0.3515	&	0.2903	&	0.6709	&	0.4371	&	0.5302	&	0.5315	&	0.8177	\\
11.2	&	0.3033	&	0.6684	&	0.5692	&	0.348	&	0.3652	&	0.7085	&	0.4489	&	0.5271	&	0.6178	&	0.8323	&	0.3033	&	0.6684	&	0.5692	&	0.348	&	0.3833	&	0.7313	&	0.4347	&	0.5275	&	0.6905	&	0.884	\\
11.3	&	0.399	&	0.7278	&	0.566	&	0.3445	&	0.48	&	0.7708	&	0.4463	&	0.5241	&	0.8018	&	0.8996	&	0.399	&	0.7278	&	0.566	&	0.3445	&	0.5024	&	0.7945	&	0.4323	&	0.5255	&	0.8932	&	0.9536	\\
11.4	&	0.5219	&	0.7906	&	0.5627	&	0.3412	&	0.6266	&	0.8358	&	0.4437	&	0.5216	&	1.036	&	0.9699	&	0.5219	&	0.7906	&	0.5627	&	0.3412	&	0.6542	&	0.8604	&	0.43	&	0.5237	&	1.15	&	1.026	\\
11.5	&	0.679	&	0.8565	&	0.5594	&	0.3379	&	0.8118	&	0.9031	&	0.4412	&	0.5199	&	1.331	&	1.043	&	0.679	&	0.8565	&	0.5594	&	0.3379	&	0.8478	&	0.9294	&	0.4278	&	0.5219	&	1.472	&	1.101	\\
11.6	&	0.8779	&	0.9251	&	0.5561	&	0.3352	&	1.047	&	0.9733	&	0.4388	&	0.5181	&	1.702	&	1.119	&	0.8779	&	0.9251	&	0.5561	&	0.3352	&	1.093	&	1.001	&	0.4255	&	0.5201	&	1.874	&	1.178	\\
11.7	&	1.128	&	0.996	&	0.553	&	0.333	&	1.343	&	1.046	&	0.4365	&	0.5163	&	2.166	&	1.197	&	1.128	&	0.996	&	0.553	&	0.333	&	1.402	&	1.076	&	0.4233	&	0.5184	&	2.373	&	1.258	\\
11.8	&	1.443	&	1.07	&	0.55	&	0.3307	&	1.715	&	1.122	&	0.4341	&	0.5145	&	2.741	&	1.277	&	1.443	&	1.07	&	0.55	&	0.3307	&	1.789	&	1.154	&	0.4211	&	0.5167	&	2.99	&	1.34	\\
11.9	&	1.837	&	1.146	&	0.547	&	0.3284	&	2.179	&	1.2	&	0.4318	&	0.5127	&	3.453	&	1.36	&	1.837	&	1.146	&	0.547	&	0.3284	&	2.272	&	1.233	&	0.419	&	0.5152	&	3.75	&	1.423	\\
12	&	2.327	&	1.225	&	0.5441	&	0.3262	&	2.755	&	1.281	&	0.4295	&	0.511	&	4.303	&	1.442	&	2.327	&	1.225	&	0.5441	&	0.3262	&	2.871	&	1.316	&	0.4168	&	0.5137	&	4.68	&	1.509	\\
12.1	&	2.934	&	1.306	&	0.5412	&	0.324	&	3.468	&	1.364	&	0.4273	&	0.5094	&	5.306	&	1.522	&	2.934	&	1.306	&	0.5412	&	0.324	&	3.61	&	1.4	&	0.4147	&	0.5123	&	5.818	&	1.596	\\
12.2	&	3.682	&	1.39	&	0.5383	&	0.3219	&	4.344	&	1.448	&	0.425	&	0.5077	&	6.522	&	1.603	&	3.682	&	1.39	&	0.5383	&	0.3219	&	4.519	&	1.486	&	0.4126	&	0.5109	&	7.203	&	1.684	\\
12.3	&	4.6	&	1.475	&	0.5355	&	0.3197	&	5.403	&	1.533	&	0.4229	&	0.5068	&	7.986	&	1.685	&	4.6	&	1.475	&	0.5355	&	0.3197	&	5.63	&	1.574	&	0.4105	&	0.5095	&	8.886	&	1.775	\\
12.4	&	5.72	&	1.562	&	0.5327	&	0.3176	&	6.671	&	1.617	&	0.4208	&	0.5068	&	9.724	&	1.766	&	5.72	&	1.562	&	0.5327	&	0.3176	&	6.983	&	1.664	&	0.4085	&	0.5081	&	10.93	&	1.866	\\
12.5	&	7.071	&	1.649	&	0.53	&	0.3161	&	8.202	&	1.702	&	0.4189	&	0.5067	&	11.8	&	1.848	&	7.082	&	1.651	&	0.5299	&	0.3155	&	8.626	&	1.754	&	0.4064	&	0.5065	&	13.39	&	1.96	\\
12.6	&	8.686	&	1.736	&	0.5275	&	0.3152	&	10.04	&	1.787	&	0.417	&	0.5065	&	14.27	&	1.929	&	8.729	&	1.74	&	0.527	&	0.3133	&	10.61	&	1.847	&	0.4043	&	0.5049	&	16.37	&	2.054	\\
12.7	&	10.62	&	1.822	&	0.5251	&	0.3143	&	12.23	&	1.871	&	0.4152	&	0.5064	&	17.18	&	2.01	&	10.71	&	1.831	&	0.5242	&	0.311	&	13	&	1.94	&	0.4021	&	0.503	&	19.96	&	2.15	\\
12.8	&	12.93	&	1.908	&	0.5228	&	0.3134	&	14.83	&	1.955	&	0.4134	&	0.5063	&	20.6	&	2.09	&	13.09	&	1.923	&	0.5213	&	0.3086	&	15.86	&	2.034	&	0.3999	&	0.5009	&	24.29	&	2.248	\\
12.9	&	15.67	&	1.993	&	0.5206	&	0.3125	&	17.9	&	2.038	&	0.4116	&	0.5059	&	24.57	&	2.168	&	15.93	&	2.015	&	0.5183	&	0.3061	&	19.29	&	2.129	&	0.3976	&	0.4985	&	29.48	&	2.347	\\
13	&	18.9	&	2.078	&	0.5183	&	0.3116	&	21.53	&	2.12	&	0.4098	&	0.5054	&	29.15	&	2.244	&	19.3	&	2.108	&	0.5153	&	0.3033	&	23.36	&	2.224	&	0.3952	&	0.4958	&	35.73	&	2.448	\\
13.1	&	22.69	&	2.161	&	0.5161	&	0.3104	&	25.76	&	2.201	&	0.4079	&	0.5049	&	34.37	&	2.316	&	23.28	&	2.2	&	0.5121	&	0.3004	&	28.19	&	2.32	&	0.3927	&	0.4928	&	43.22	&	2.551	\\
13.2	&	27.12	&	2.242	&	0.5139	&	0.3092	&	30.67	&	2.279	&	0.406	&	0.5044	&	40.27	&	2.384	&	27.97	&	2.293	&	0.5089	&	0.2972	&	33.9	&	2.417	&	0.3901	&	0.4893	&	52.21	&	2.656	\\
13.3	&	32.27	&	2.322	&	0.5116	&	0.3078	&	36.33	&	2.354	&	0.404	&	0.5039	&	46.99	&	2.449	&	33.46	&	2.385	&	0.5054	&	0.2937	&	40.62	&	2.513	&	0.3873	&	0.4853	&	62.98	&	2.762	\\
13.4	&	38.2	&	2.399	&	0.5093	&	0.3066	&	42.8	&	2.426	&	0.402	&	0.5038	&	54.61	&	2.511	&	39.88	&	2.477	&	0.5018	&	0.29	&	48.52	&	2.61	&	0.3844	&	0.481	&	75.88	&	2.87	\\
13.5	&	45.05	&	2.472	&	0.5069	&	0.3056	&	50.18	&	2.494	&	0.4001	&	0.5042	&	63.25	&	2.569	&	47.37	&	2.569	&	0.4981	&	0.2862	&	57.81	&	2.708	&	0.3813	&	0.4765	&	91.32	&	2.981	\\
13.6	&	52.85	&	2.542	&	0.5046	&	0.3048	&	58.56	&	2.557	&	0.3982	&	0.5045	&	72.97	&	2.624	&	56.06	&	2.66	&	0.4942	&	0.2821	&	68.68	&	2.806	&	0.3782	&	0.4714	&	109.8	&	3.093	\\
13.7	&	61.72	&	2.608	&	0.5024	&	0.3044	&	68.09	&	2.618	&	0.3963	&	0.5049	&	83.87	&	2.675	&	66.14	&	2.752	&	0.4901	&	0.2779	&	81.38	&	2.905	&	0.375	&	0.466	&	131.9	&	3.208	\\
13.8	&	71.81	&	2.67	&	0.5002	&	0.3042	&	78.88	&	2.674	&	0.3945	&	0.5055	&	96.04	&	2.722	&	77.81	&	2.842	&	0.4859	&	0.2735	&	96.2	&	3.004	&	0.3716	&	0.4604	&	158.4	&	3.326	\\
13.9	&	83.26	&	2.728	&	0.4982	&	0.3041	&	91.07	&	2.727	&	0.3928	&	0.5062	&	109.6	&	2.765	&	91.29	&	2.933	&	0.4816	&	0.2689	&	113.5	&	3.104	&	0.3682	&	0.4544	&	190	&	3.446	\\
14	&	96.2	&	2.781	&	0.4962	&	0.3042	&	104.8	&	2.775	&	0.3911	&	0.507	&	124.6	&	2.804	&	106.8	&	3.024	&	0.4772	&	0.2642	&	133.6	&	3.204	&	0.3647	&	0.4481	&	227.8	&	3.568	\\
14.1	&	110.8	&	2.831	&	0.4944	&	0.3043	&	120.1	&	2.819	&	0.3895	&	0.5079	&	141.3	&	2.839	&	124.7	&	3.114	&	0.4727	&	0.2593	&	157	&	3.306	&	0.3612	&	0.4415	&	273	&	3.694	\\
14.2	&	127.2	&	2.876	&	0.4926	&	0.3045	&	137.3	&	2.859	&	0.388	&	0.509	&	159.6	&	2.87	&	145.2	&	3.205	&	0.468	&	0.2542	&	184.1	&	3.408	&	0.3575	&	0.4345	&	327.1	&	3.823	\\
14.3	&	145.5	&	2.917	&	0.491	&	0.3049	&	156.4	&	2.894	&	0.3865	&	0.5102	&	179.9	&	2.897	&	168.7	&	3.295	&	0.4633	&	0.2489	&	215.5	&	3.511	&	0.3538	&	0.4271	&	391.8	&	3.955	\\
14.4	&	166	&	2.954	&	0.4894	&	0.3054	&	177.5	&	2.925	&	0.3851	&	0.5115	&	202.3	&	2.92	&	195.5	&	3.386	&	0.4584	&	0.2435	&	251.8	&	3.615	&	0.3499	&	0.4195	&	469.1	&	4.091	\\

\hline
\end{tabular}
}

\end{sidewaystable}

\subsection{Tabulating the Ternary Diagram}

For the $3$-layer model of solid exoplanet, points of a curve segment on the ternary diagram represent all the solutions for a given mass-radius input. These ternary diagrams are tabulated (Table~\ref{Table3}) with the intent to make comparison to observations easier. 

Usually, there are infinite combinations (solutions) of Fe, MgSiO${}_{3}$ and H${}_{2}$O mass fractions which can give the same mass-radius pair. All the combinations together form a curve segment on the ternary diagram of Fe, MgSiO${}_{3}$ and H${}_{2}$O mass fractions~\citep{Zeng_Seager:2008, Valencia:2007b}. This curve segment can be approximated by $3$ points on it: two endpoints where one or more out of the $3$ layers are absent and one point in between where all $3$ layers are present to give the same mass and radius. The two endpoints correspond to the minimum central pressure (p0${}_{min}$) and maximum central pressure (p0${}_{max}$) allowed for the given mass-radius pair. The middle point is chosen to have the central pressure p0${}_{mid}$=$\sqrt{\text{p0}{}_{max}*\text{p0}{}_{min}}$. 

Table~\ref{Table3} contains 8 columns:

1st column: $Mass$. The masses range from 0.1 $M_{\oplus}$ to 100 $M_{\oplus}$ with 41 points in total. The range between 0.1 and 1 $M_{\oplus}$ is equally divided into 10 sections in logarithmic scale. The range between 1 and 10 $M_{\oplus}$ is equally divided into 20 sections in logarithmic scale. And the range between 10 and 100 $M_{\oplus}$ is equally divided into 10 sections in logarithmic scale. 

2nd column: $Radius$. For each mass $M$ in Table~\ref{Table3} there are $12$ radius values, $11$ of which are equally spaced within the allowed range: $R_{Fe}(M)+(R_{H_2O}(M)-R_{Fe}(M))*i$, where $i=0, 0.1, 0.2,... , 0.9, 1.0$. The $12$-th radius value ($R_{MgSiO_3}(M)$) is inserted into the list corresponding to the pure-MgSiO${}_{3}$ planet radius (see Table~\ref{Table1}) for mass $M$. Here $R_{Fe}(M)$, $R_{MgSiO_3}(M)$, and $R_{H_2O}(M)$ are the radii for planets with mass $M$ composed of pure-Fe, pure-MgSiO${}_{3}$, and pure-H${}_{2}$O correspondingly. 

Overall, there are $41*12=492$ different mass-radius pairs in Table~\ref{Table3}.
For each $(M,R)$, $3$ cases: p0${}_{min}$, p0${}_{mid}$, and p0${}_{max}$ are listed. 

3rd column: central pressure p0 (Pascal) in logarithmic base-10 scale.

4th column: p1/p0, the ratio of p1 (the first boundary pressure, i.e., the pressure at the Fe-MgSiO${}_{3}$ boundary) over p0.

5th column: p2/p1, the ratio of p2 (the second boundary pressure, i.e., the pressure at the MgSiO${}_{3}$-H${}_{2}$O boundary) over p1.

6th column: Fe mass fraction (the ratio of the Fe-layer mass over the total mass of the planet).

7th column: MgSiO${}_{3}$ mass fraction (the ratio of the MgSiO${}_{3}$-layer mass over the total mass of the planet).

8th column: H${}_{2}$O mass fraction (the ratio of the H${}_{2}$O-layer mass of over the total mass of the planet).

6th, 7th and 8th columns always add up to one. 
\\

\begin{center}
\begin{longtable}{|c|c|c|c|c|c|c|c|}

\caption{Table for Ternary Diagram} \label{Table3}\\

\hline \multicolumn{1}{|c|}{\textbf{M($M_{\oplus}$)}} & 
\multicolumn{1}{c|}{\textbf{R($R_{\oplus}$)}} & 
\multicolumn{1}{c|}{$\mathbf{log_{10}(p0)}$} & 
\multicolumn{1}{c|}{\textbf{p1/p0}} & 
\multicolumn{1}{c|}{\textbf{p2/p1}} & 
\multicolumn{1}{c|}{$\mathbf{Fe}$} & 
\multicolumn{1}{c|}{$\mathbf{MgSiO_3}$} & 
\multicolumn{1}{c|}{$\mathbf{H_2O}$} \\ \hline
\endfirsthead

\multicolumn{8}{c}%
{{\bfseries \tablename\ \thetable{} -- continued from previous page}} \\
\hline \multicolumn{1}{|c|}{\textbf{M($M_{\oplus}$)}} & 
\multicolumn{1}{c|}{\textbf{R($R_{\oplus}$)}} & 
\multicolumn{1}{c|}{$\mathbf{log_{10}(p0)}$} & 
\multicolumn{1}{c|}{\textbf{p1/p0}} & 
\multicolumn{1}{c|}{\textbf{p2/p1}} & 
\multicolumn{1}{c|}{$\mathbf{Fe}$} & 
\multicolumn{1}{c|}{$\mathbf{MgSiO_3}$} & 
\multicolumn{1}{c|}{$\mathbf{H_2O}$} \\ \hline
\endhead

\hline \multicolumn{8}{|r|}{{Continued on next page}} \\ \hline
\endfoot
 
 \hline \hline
\endlastfoot

0.1	&	0.3888	&	10.9212	&	0	&	0	&	1	&	0	&	0	\\
0.1	&	0.3888	&	10.9212	&	0	&	0	&	1	&	0	&	0	\\
0.1	&	0.3888	&	10.9212	&	0	&	0	&	1	&	0	&	0	\\
0.1	&	0.4226	&	10.9066	&	0.133	&	0	&	0.759	&	0.241	&	0	\\
0.1	&	0.4226	&	10.9126	&	0.102	&	0.046	&	0.818	&	0.172	&	0.01	\\
0.1	&	0.4226	&	10.9186	&	0.024	&	1	&	0.953	&	0	&	0.047	\\
\vdots & \vdots & \vdots & \vdots & \vdots & \vdots & \vdots & \vdots \\
100	&	4.102	&	13.0979	&	1	&	1	&	0	&	0	&	1	\\

\end{longtable}
\end{center}

A dynamic and interactive tool to characterize and illustrate the interior structure of exoplanets built upon Table~\ref{Table3} and other models in this paper is available on the website \url{http://www.cfa.harvard.edu/~lzeng}. 

\subsection{Generate curve segment on ternary diagram using Table~\ref{Table3}}
\begin{figure}[htbp]
\begin{center}
\includegraphics[scale=0.6,angle=90]{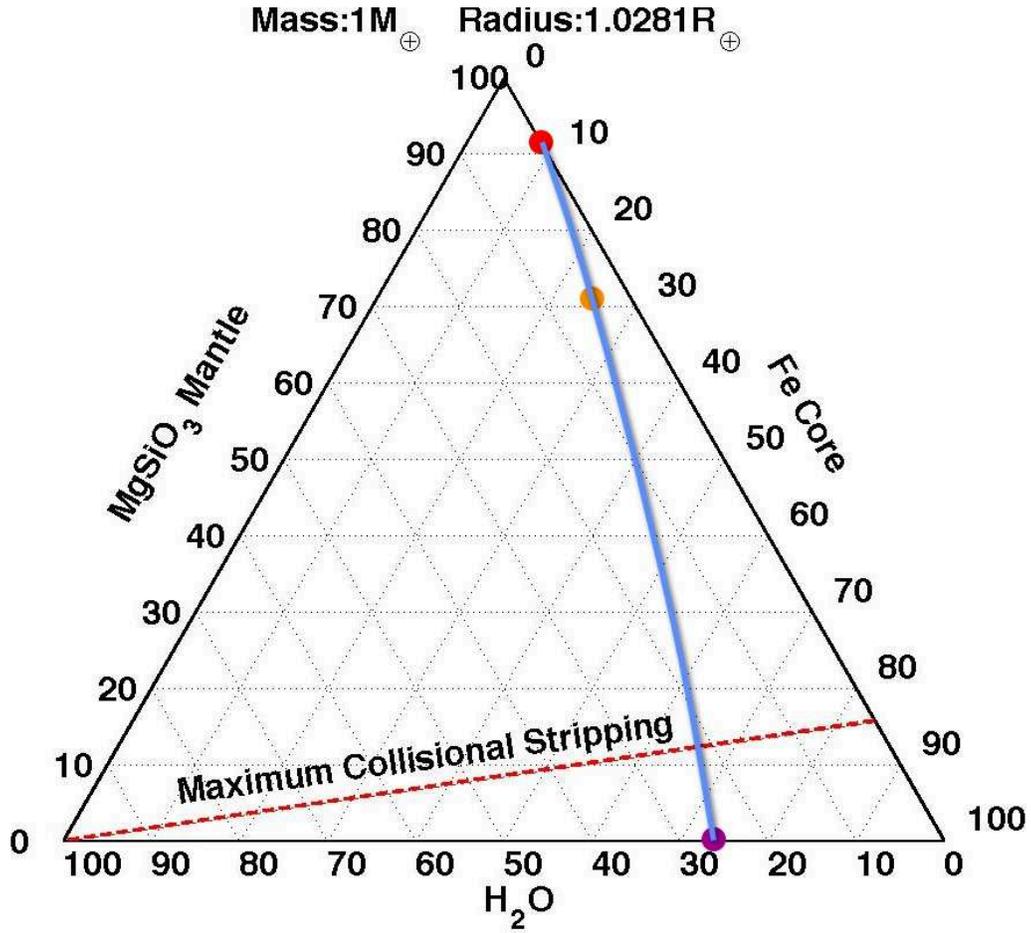}
\end{center}
\caption{The red, orange and purple points correspond to $log_{10}(p0~(in~Pa))=11.4391$, $11.5863$, and $11.7336$. The mass fractions are (1) red point: FeMF=0.083, MgSiO${}_{3}$MF=0.917, H${}_{2}$OMF=0; (2) orange point: FeMF=0.244, MgSiO${}_{3}$MF=0.711, H${}_{2}$OMF=0.046; (3) purple point: FeMF=0.728, MgSiO${}_{3}$MF=0, H${}_{2}$OMF=0.272. MF here stands for mass fraction. The blue curve is the parabolic fit. The red dashed curve is the maximum collisional stripping curve by~\citet{Marcus:2010}.}
\label{ternary}
\end{figure}

One utility of Table~\ref{Table3} is to generate the curve segment on the $3$-layer ternary diagram for a given mass-radius pair. As an example, for M=$1M_{\oplus}$ and R=$1.0281R_{\oplus}$, the table provides $3$ p0's. For each p0, the mass fractions of Fe, MgSiO${}_{3}$, and H${}_{2}$O are given to determine a point on the ternary diagram. Then, a parabolic fit (see Fig.~\ref{ternary}) through the $3$ points is a good approximation to the actual curve segment. This parabola may intersect the maximum collisional stripping curve by~\citet{Marcus:2010}, indicating that the portion of parabola beneath the intersection point may be ruled out by planet formation theory.  

\section{Conclusion}
The $2$-layer and $3$-layer models for solid exoplanets composed of Fe, MgSiO${}_{3}$, and H${}_{2}$O are the focus of this paper. The mass-radius contours (Fig.~\ref{contourplots}) are provided for the $2$-layer model, useful for readers to quickly calculate the interior structure of a solid exoplanet. 
The $2$-parameter contour mesh may also help one build physical insights into the solid exoplanet interior structure. 

The complete $3$-layer mass-fraction ternary diagram is tabulated (Table~\ref{Table3}), useful for readers to interpolate and calculate all solutions as the mass fractions of the $3$ layers for a given mass-radius input. The details of the EOS of Fe, MgSiO${}_{3}$, and H${}_{2}$O and how they are calculated and used in this paper are discussed in section~\ref{EOS} and shown in Fig.~\ref{eosplot}. 

A dynamic and interactive tool to characterize and illustrate the interior structure of exoplanets built upon Table~\ref{Table3} and other models in this paper is available on the website \url{http://www.cfa.harvard.edu/~lzeng}. 

The effect of Fe partitioning between mantle and core on mass-radius relation is explored in section~\ref{nondiff}, and the result is shown in Fig.~\ref{f4} and Table~\ref{Table2}. 

With the ongoing Kepler Mission and many other exoplanet searching projects, we hope this paper could provide a handy tool for observers to fast characterize the interior structure of exoplanets already discovered or to be discovered, and further our understanding of those worlds beyond our own. 

\section{Acknowledgements}
We acknowledge partial support for this work by NASA co-operative agreement NNX09AJ50A  (Kepler mission science team).

We would like to thank Michail Petaev and Stein Jacobsen for their valuable comments and suggestions. This research is supported by the National Nuclear Security Administration under the High Energy Density Laboratory Plasmas through DOE grant $\#$ DE-FG52-09NA29549 to S. B. Jacobsen (PI) with Harvard University. This research is the authors' views and not those of the DOE. 

Li Zeng would like to thank Professor Pingyuan Li, Li Zeng's grandfather, in the Department of Mathematics at Chongqing University, for giving Li Zeng important spiritual support and guidance on research. The guidance includes research strategy and approach, methods of solving differential equations and other numeric methods, etc. 

Li Zeng would also like to give special thanks to Master Anlin Wang. Master Wang is a Traditional Chinese Kung Fu Master and World Champion. He is also a practitioner and realizer of Traditional Chinese Philosophy of Tao Te Ching, which is the ancient oriental wisdom to study the relation between the universe, nature and humanity. Valuable inspirations were obtained through discussion of Tao Te Ching with Master Wang as well as Qigong cultivation with him. 

\clearpage


\bibliographystyle{bibstyle1}
\bibliography{mybib}

\begin{thebibliography}{57}
\expandafter\ifx\csname natexlab\endcsname\relax\def\natexlab#1{#1}\fi

\bibitem[{{Anderson} {et~al.}(2001){Anderson}, {Dubrovinsky}, {Saxena}, \&
  {LeBihan}}]{Anderson:2001}
{Anderson}, O.~L., {Dubrovinsky}, L., {Saxena}, S.~K., \& {LeBihan}, T. 2001,
  \grl, 28, 399

\bibitem[{Asplund {et~al.}(2009)Asplund, Grevesse, Sauval, \&
  Scott}]{Asplund:2009}
Asplund, M., Grevesse, N., Sauval, A.~J., \& Scott, P. 2009, Annual Review of
  Astronomy and Astrophysics, 47, 481

\bibitem[{Batalha {et~al.}(2011)Batalha, Borucki, Bryson, Buchhave, Caldwell,
  Christensen-Dalsgaard, Ciardi, Dunham, Fressin, III, Gilliland, Haas, Howell,
  Jenkins, Kjeldsen, Koch, Latham, Lissauer, Marcy, Rowe, Sasselov, Seager,
  Steffen, Torres, Basri, Brown, Charbonneau, Christiansen, Clarke, Cochran,
  Dupree, Fabrycky, Fischer, Ford, Fortney, Girouard, Holman, Johnson,
  Isaacson, Klaus, Machalek, Moorehead, Morehead, Ragozzine, Tenenbaum,
  Twicken, Quinn, VanCleve, Walkowicz, Welsh, Devore, \& Gould}]{Batalha:2011}
Batalha, N.~M., Borucki, W.~J., Bryson, S.~T., {et~al.} 2011, The Astrophysical
  Journal, 729, 27

\bibitem[{Bina(2003)}]{Bina:2003}
Bina, C. 2003, in Treatise on Geochemistry, ed. E.~in~Chief:å å Heinrich
  D.~Holland \& K.~K. Turekian (Oxford: Pergamon), 39 -- 59

\bibitem[{{Birch}(1947)}]{Birch:1947}
{Birch}, F. 1947, Physical Review, 71, 809

\bibitem[{Borucki {et~al.}(2012)Borucki, Koch, Batalha, Bryson, Rowe, Fressin,
  Torres, Caldwell, Christensen-Dalsgaard, Cochran, DeVore, Gautier, Geary,
  Gilliland, Gould, Howell, Jenkins, Latham, Lissauer, Marcy, Sasselov, Boss,
  Charbonneau, Ciardi, Kaltenegger, Doyle, Dupree, Ford, Fortney, Holman,
  Steffen, Mullally, Still, Tarter, Ballard, Buchhave, Carter, Christiansen,
  Demory, Désert, Dressing, Endl, Fabrycky, Fischer, Haas, Henze, Horch,
  Howard, Isaacson, Kjeldsen, Johnson, Klaus, Kolodziejczak, Barclay, Li,
  Meibom, Prsa, Quinn, Quintana, Robertson, Sherry, Shporer, Tenenbaum,
  Thompson, Twicken, Cleve, Welsh, Basu, Chaplin, Miglio, Kawaler, Arentoft,
  Stello, Metcalfe, Verner, Karoff, Lundkvist, Lund, Handberg, Elsworth,
  Hekker, Huber, Bedding, \& Rapin}]{Borucki:2012}
Borucki, W.~J., Koch, D.~G., Batalha, N., {et~al.} 2012, The Astrophysical
  Journal, 745, 120

\bibitem[{{Caracas} \& {Cohen}(2008)}]{Caracas:2008}
{Caracas}, R., \& {Cohen}, R.~E. 2008, Physics of the Earth and Planetary
  Interiors, 168, 147

\bibitem[{Carter {et~al.}(2012)Carter, Agol, Chaplin, Basu, Bedding, Buchhave,
  Christensen-Dalsgaard, Deck, Elsworth, Fabrycky, Ford, Fortney, Hale,
  Handberg, Hekker, Holman, Huber, Karoff, Kawaler, Kjeldsen, Lissauer, Lopez,
  Lund, Lundkvist, Metcalfe, Miglio, Rogers, Stello, Borucki, Bryson,
  Christiansen, Cochran, Geary, Gilliland, Haas, Hall, Howard, Jenkins, Klaus,
  Koch, Latham, MacQueen, Sasselov, Steffen, Twicken, \& Winn}]{Carter:2012}
Carter, J.~A., Agol, E., Chaplin, W.~J., {et~al.} 2012, Science, 337, 556

\bibitem[{Chaplin(2012)}]{Chaplin:2012}
Chaplin, M. 2012, Water Phase Diagram,
  \url{http://www.lsbu.ac.uk/water/phase.html#bb}

\bibitem[{{Charbonneau} {et~al.}(2009){Charbonneau}, {Berta}, {Irwin}, {Burke},
  {Nutzman}, {Buchhave}, {Lovis}, {Bonfils}, {Latham}, {Udry}, {Murray-Clay},
  {Holman}, {Falco}, {Winn}, {Queloz}, {Pepe}, {Mayor}, {Delfosse}, \&
  {Forveille}}]{Charbonneau:2009}
{Charbonneau}, D., {Berta}, Z.~K., {Irwin}, J., {et~al.} 2009, \nat, 462, 891

\bibitem[{Choukroun \& Grasset(2007)}]{Choukroun:2007}
Choukroun, M., \& Grasset, O. 2007, The Journal of Chemical Physics, 127,
  124506

\bibitem[{Cochran {et~al.}(2011)Cochran, Fabrycky, Torres, Fressin, Désert,
  Ragozzine, Sasselov, Fortney, Rowe, Brugamyer, Bryson, Carter, Ciardi,
  Howell, Steffen, Borucki, Koch, Winn, Welsh, Uddin, Tenenbaum, Still, Seager,
  Quinn, Mullally, Miller, Marcy, MacQueen, Lucas, Lissauer, Latham, Knutson,
  Kinemuchi, Johnson, Jenkins, Isaacson, Howard, Horch, Holman, Henze, Haas,
  Gilliland, III, Ford, Fischer, Everett, Endl, Demory, Deming, Charbonneau,
  Caldwell, Buchhave, Brown, \& Batalha}]{Cochran:2011}
Cochran, W.~D., Fabrycky, D.~C., Torres, G., {et~al.} 2011, The Astrophysical
  Journal Supplement Series, 197, 7

\bibitem[{Daucik \& Dooley(2011)}]{IAPWS:2011}
Daucik, K., \& Dooley, R.~B. 2011, Revised Release on the Pressure along the
  Melting and Sublimation Curves of Ordinary Water Substance (The International
  Association for the Properties of Water and Steam)

\bibitem[{Dunaeva {et~al.}(2010)Dunaeva, Antsyshkin, \& Kuskov}]{Dunaeva:2010}
Dunaeva, A.~N., Antsyshkin, D.~V., \& Kuskov, O.~L. 2010, Solar System
  Research, 44, 202

\bibitem[{Eliezer {et~al.}(2002)Eliezer, Ghatak, \& Hora}]{Eliezer:2002}
Eliezer, S., Ghatak, A., \& Hora, H. 2002, Fundamentals of Equations of State
  (London: World Scientific)

\bibitem[{Fei {et~al.}(2007)Fei, Zhang, Corgne, Watson, Ricolleau, Meng, \&
  Prakapenka}]{Fei:2007}
Fei, Y., Zhang, L., Corgne, A., {et~al.} 2007, Geophys. Res. Lett., 34, L17307

\bibitem[{Fortney {et~al.}(2007)Fortney, Marley, \& Barnes}]{Fortney:2007}
Fortney, J., Marley, M., \& Barnes, J. 2007, The Astrophysical Journal, 659,
  1661

\bibitem[{Frank {et~al.}(2004)Frank, Fei, \& Hu}]{Frank:2004}
Frank, M.~R., Fei, Y., \& Hu, J. 2004, Geochimica et Cosmochimica Acta, 68,
  2781

\bibitem[{French {et~al.}(2009)French, Mattsson, Nettelmann, \&
  Redmer}]{French:2009}
French, M., Mattsson, T.~R., Nettelmann, N., \& Redmer, R. 2009, Phys. Rev. B,
  79, 054107

\bibitem[{{Fressin} {et~al.}(2012){Fressin}, {Torres}, {Rowe}, {Charbonneau},
  {Rogers}, {Ballard}, {Batalha}, {Borucki}, {Bryson}, {Buchhave}, {Ciardi},
  {D{\'e}sert}, {Dressing}, {Fabrycky}, {Ford}, {Gautier}, {Henze}, {Holman},
  {Howard}, {Howell}, {Jenkins}, {Koch}, {Latham}, {Lissauer}, {Marcy},
  {Quinn}, {Ragozzine}, {Sasselov}, {Seager}, {Barclay}, {Mullally}, {Seader},
  {Still}, {Twicken}, {Thompson}, \& {Uddin}}]{Fressin:2011}
{Fressin}, F., {Torres}, G., {Rowe}, J.~F., {et~al.} 2012, \nat, 482, 195

\bibitem[{Goncharov {et~al.}(2005)Goncharov, Goldman, Fried, Crowhurst, Kuo,
  Mundy, \& Zaug}]{Goncharov:2005}
Goncharov, A.~F., Goldman, N., Fried, L.~E., {et~al.} 2005, Phys. Rev. Lett.,
  94, 125508

\bibitem[{Grasset {et~al.}(2009)Grasset, Schneider, \& Sotin}]{Grasset:2009}
Grasset, O., Schneider, J., \& Sotin, C. 2009, The Astrophysical Journal, 693,
  722

\bibitem[{Hatzes {et~al.}(2011)Hatzes, Fridlund, Nachmani, Mazeh, Valencia,
  Hébrard, Carone, Pätzold, Udry, Bouchy, Deleuil, Moutou, Barge, Bordé,
  Deeg, Tingley, Dvorak, Gandolfi, Ferraz-Mello, Wuchterl, Guenther, Guillot,
  Rauer, Erikson, Cabrera, Csizmadia, Léger, Lammer, Weingrill, Queloz,
  Alonso, Rouan, \& Schneider}]{Hatzes:2011}
Hatzes, A.~P., Fridlund, M., Nachmani, G., {et~al.} 2011, The Astrophysical
  Journal, 743, 75

\bibitem[{Hirose(2010)}]{Hirose:2010}
Hirose, K. 2010, Scientific American, 302, 76

\bibitem[{Hirsch \& Holzapfel(1984)}]{Hirsch:1984}
Hirsch, K., \& Holzapfel, W. 1984, Journal of Experimental and Theoretical
  Physics, 101, 142

\bibitem[{Howell {et~al.}(2012)Howell, Rowe, Bryson, Quinn, Marcy, Isaacson,
  Ciardi, Chaplin, Metcalfe, Monteiro, Appourchaux, Basu, Creevey, Gilliland,
  Quirion, Stello, Kjeldsen, Christensen-Dalsgaard, Elsworth, García, Houdek,
  Karoff, Molenda-Żakowicz, Thompson, Verner, Torres, Fressin, Crepp, Adams,
  Dupree, Sasselov, Dressing, Borucki, Koch, Lissauer, Latham, Buchhave, III,
  Everett, Horch, Batalha, Dunham, Szkody, Silva, Mighell, Holberg, Ballot,
  Bedding, Bruntt, Campante, Handberg, Hekker, Huber, Mathur, Mosser, Régulo,
  White, Christiansen, Middour, Haas, Hall, Jenkins, McCaulif, Fanelli, Kulesa,
  McCarthy, \& Henze}]{Howell:2012}
Howell, S.~B., Rowe, J.~F., Bryson, S.~T., {et~al.} 2012, The Astrophysical
  Journal, 746, 123

\bibitem[{III {et~al.}(2012)III, Charbonneau, Rowe, Marcy, Isaacson, Torres,
  Fressin, Rogers, Desert, Buchhave, Latham, Quinn, Ciardi, Fabrycky, Ford,
  Gilliland, Walkowicz, Bryson, Cochran, Endl, Fischer, Howell, Horch, Barclay,
  Batalha, Borucki, Christiansen, Geary, Henze, Holman, Ibrahim, Jenkins,
  Kinemuchi, Koch, Lissauer, Sanderfer, Sasselov, Seager, Silverio, Smith,
  Still, Stumpe, Tenenbaum, \& Cleve}]{Gautier:2012}
III, T. N.~G., Charbonneau, D., Rowe, J.~F., {et~al.} 2012, The Astrophysical
  Journal, 749, 15

\bibitem[{Karki {et~al.}(2000)Karki, Wentzcovitch, de~Gironcoli, \&
  Baroni}]{Karki:2000}
Karki, B.~B., Wentzcovitch, R.~M., de~Gironcoli, S., \& Baroni, S. 2000, Phys.
  Rev. B, 62, 14750

\bibitem[{Kresse \& Furthm{\"u}ller(1996)}]{Kresse:1996}
Kresse, G., \& Furthm{\"u}ller, J. 1996, Physical Review B, 54, 11169

\bibitem[{Kresse \& Hafner(1993)}]{Kresse:1993}
Kresse, G., \& Hafner, J. 1993, Physical Review B, 47, 558

\bibitem[{Kresse \& Hafner(1994)}]{Kresse:1994}
---. 1994, Journal of Physics: Condensed Matter, 6, 8245

\bibitem[{Leger {et~al.}(2009)Leger, Rouan, Schneider, Barge, Fridlund, \&
  et~al.}]{Leger:2009}
Leger, A., Rouan, D., Schneider, J., {et~al.} 2009, Astronomy and Astrophysics,
  506, 287

\bibitem[{Lissauer {et~al.}(2011)Lissauer, Fabrycky, Ford, Borucki, Fressin, \&
  et~al.}]{Lissauer:2011}
Lissauer, J.~J., Fabrycky, D.~C., Ford, E.~B., {et~al.} 2011, Nature, 470, 53

\bibitem[{{Macfarlane}(1984)}]{Macfarlane:1984}
{Macfarlane}, J.~J. 1984, \apj, 280, 339

\bibitem[{Marcus {et~al.}(2010)Marcus, Sasselov, Hernquist, \&
  Stewart}]{Marcus:2010}
Marcus, R., Sasselov, D., Hernquist, L., \& Stewart, S. 2010, The Astrophysical
  Journal Letters, 712, L73

\bibitem[{Murakami {et~al.}(2004)Murakami, Hirose, Kawamura, Sata, \&
  Ohishi}]{Murakami:2004}
Murakami, M., Hirose, K., Kawamura, K., Sata, N., \& Ohishi, Y. 2004, Science,
  304, 855

\bibitem[{Oganov \& Ono(2004)}]{Oganov:2004}
Oganov, A.~R., \& Ono, S. 2004, Nature, 430, 445

\bibitem[{{Poirier}(2000)}]{Poirier:2000}
{Poirier}, J.-P. 2000, {Introduction to the Physics of the Earth's Interior}
  (Cambridge, UK: Cambridge University Press)

\bibitem[{Queloz {et~al.}(2009)Queloz, Bouchy, Moutou, Hatzes, \&
  H{\'e}brard}]{Queloz:2009}
Queloz, D., Bouchy, F., Moutou, C., Hatzes, A., \& H{\'e}brard, G. 2009,
  Astronomy and Astrophysics, 506, 303

\bibitem[{Robinson {et~al.}(1996)Robinson, Zhu, Singh, \&
  Evans}]{Robinson:1996}
Robinson, G.~W., Zhu, S.~B., Singh, S., \& Evans, M.~W. 1996, Water in Biology,
  Chemistry and Physics: Experimental Overviews and Computational Methodologies
  (Singapore: World Scientific)

\bibitem[{Salpeter \& Zapolsky(1967)}]{Salpeter:1967}
Salpeter, E.~E., \& Zapolsky, H.~S. 1967, Physical Review, 158, 876

\bibitem[{Seager {et~al.}(2007)Seager, Kuchner, Hier-Majumder, \&
  Militzer}]{Seager:2007}
Seager, S., Kuchner, M., Hier-Majumder, C.~A., \& Militzer, B. 2007, The
  Astrophysical Journal, 669, 1279

\bibitem[{Sotin {et~al.}(2007)Sotin, Grasset, \& Mocquet}]{Sotin:2007}
Sotin, C., Grasset, O., \& Mocquet, A. 2007, Icarus, 191, 337

\bibitem[{Sotin {et~al.}(2010)Sotin, Jackson, \& Seager}]{Seager:2010}
Sotin, C., Jackson, J.~M., \& Seager, S. 2010, Exoplanets (The University of
  Arizona Press), 375--395

\bibitem[{Spera {et~al.}(2006)Spera, Yuen, \& Giles}]{Spera:2006}
Spera, F.~J., Yuen, D.~A., \& Giles, G. 2006, Physics of the Earth and
  Planetary Interiors, 159, 234

\bibitem[{Stixrude \& Lithgow-Bertelloni(2011)}]{Stixrude:2011}
Stixrude, L., \& Lithgow-Bertelloni, C. 2011, Geophysical Journal
  International, 184, 1180

\bibitem[{Umemoto \& Wentzcovitch(2011)}]{Umemoto:2011}
Umemoto, K., \& Wentzcovitch, R.~M. 2011, Earth and Planetary Science Letters,
  311, 225

\bibitem[{Valencia {et~al.}(2010)Valencia, Ikoma, Guillot, \&
  Nettelmann}]{Valencia:2010}
Valencia, D., Ikoma, M., Guillot, T., \& Nettelmann, N. 2010, Astronomy and
  Astrophysics, 516

\bibitem[{Valencia {et~al.}(2006)Valencia, O'Connell, \&
  Sasselov}]{Valencia:2006}
Valencia, D., O'Connell, R.~J., \& Sasselov, D. 2006, Icarus, 181, 545

\bibitem[{Valencia {et~al.}(2007{\natexlab{a}})Valencia, Sasselov, \&
  O'Connell}]{Valencia:2007b}
Valencia, D., Sasselov, D.~D., \& O'Connell, R.~J. 2007{\natexlab{a}}, The
  Astrophysical Journal, 665, 1413

\bibitem[{Valencia {et~al.}(2007{\natexlab{b}})Valencia, Sasselov, \&
  O'Connell}]{Valencia:2007a}
---. 2007{\natexlab{b}}, The Astrophysical Journal, 656, 545

\bibitem[{Vinet {et~al.}(1987)Vinet, Ferrante, Rose, \& Smith}]{Vinet:1987}
Vinet, P., Ferrante, J., Rose, J.~H., \& Smith, J.~R. 1987, Journal of
  Geophysical Research, 92, 9319

\bibitem[{Vinet {et~al.}(1989)Vinet, Rose, Ferrante, \& Smith}]{Vinet:1989}
Vinet, P., Rose, J.~H., Ferrante, J., \& Smith, J.~R. 1989, Journal of Physics:
  Condensed Matter, 1, 1941

\bibitem[{Winn {et~al.}(2011)Winn, Matthews, Dawson, Fabrycky, \&
  Holman}]{Winn:2011}
Winn, J.~N., Matthews, J.~M., Dawson, R.~I., Fabrycky, D., \& Holman, M.~J.
  2011, The Astrophysical Journal Letters, 737, L18(6pp)

\bibitem[{Wu {et~al.}(2011)Wu, Umemoto, Ji, Wang, Ho, \&
  Wentzcovitch}]{Wu:2011}
Wu, S., Umemoto, K., Ji, M., {et~al.} 2011, Phys. Rev. B, 83, 184102

\bibitem[{Yu \& Jacobsen(2011)}]{Yu:2011}
Yu, G., \& Jacobsen, S.~B. 2011, PNAS, 108, 17604

\bibitem[{Zeng \& Seager(2008)}]{Zeng_Seager:2008}
Zeng, L., \& Seager, S. 2008, Publications of the Astronomical Society of the
  Pacific, 120, 983

\end{thebibliography}

\end{document}